%% file: paper.tex
\documentclass[sigconf]{acmart}

\renewcommand\footnotetextcopyrightpermission[1]{} 
\AtBeginDocument{%
  \providecommand\BibTeX{{%
    \normalfont B\kern-0.5em{\scshape i\kern-0.25em b}\kern-0.8em\TeX}}}

\usepackage{multirow}
\usepackage{multicol}
\usepackage{color}
\usepackage{soul}
\usepackage{booktabs}
\usepackage{url}
\usepackage{subcaption}
\usepackage{colortbl}
\usepackage{pdflscape}
\usepackage{rotating}
\usepackage{tabularx}
\usepackage{xspace}
\usepackage{adjustbox}

\begin{document}

\title{An Empirical Analysis of HTTPS Configuration Security}


\author{Camelia Simoiu}
\authornote{Both authors contributed equally to this research.}
\affiliation{%
 \institution{Stanford University}}
\email{csimoiu@stanford.edu}

\author{Wilson Nguyen}
\affiliation{%
 \institution{Stanford University}}
\email{wdnguyen@cs.stanford.edu}
\authornotemark[1]

\author{Zakir Durumeric}
\affiliation{%
 \institution{Stanford University}}
\email{zakir@cs.stanford.edu}


\newcommand{\ts}{\setlength{\tabcolsep}{-2pt}}
\newcommand{\tl}{\setlength{\tabcolsep}{6pt}}
\newcommand{\TK}{\hl{\bf TK}\xspace}
\newcommand{\tk}{\TK}

\renewcommand{\paragraph}[1]{\vspace{1pt}\noindent\textbf{#1.\enspace}}

\begin{abstract}

It is notoriously difficult to securely configure HTTPS, and poor server configurations have contributed to several attacks including the FREAK, Logjam, and POODLE attacks.
In this work, we empirically evaluate the TLS security posture of popular websites and endeavor to understand the configuration decisions that operators
make. We correlate several sources of influence on sites’ security
postures, including software defaults, cloud providers, and online
recommendations.
We find a fragmented web ecosystem: while most websites have secure configurations, this is largely due to major cloud providers that offer secure defaults.
Individually configured servers are more often insecure than not.
This may be in part because common resources available to individual operators---server software defaults and online configuration guides---are frequently insecure.
Our findings highlight the importance of considering SaaS services separately from individually-configured sites in measurement studies, and the need for server software to ship with secure defaults.
\end{abstract}


\settopmatter{printfolios=true}
\maketitle
\pagestyle{plain}

\newcommand{\wilson}[1]{{\textcolor{violet}{[\textit{#1} --WN]}}}
\newcommand{\zakir}[1]{{\textcolor{red}{[\textit{#1} --ZD]}}}
\newcommand{\camelia}[1]{{\textcolor{blue}{[\textit{#1} --CS]}}}

\input{1_introduction}

\input{9_related}

\input{2_background}

\input{3_methodology}

\input{4_ecosystem}

\input{5_model}
\input{6_defaults}

\input{7_recommendations}

\input{8_discussion}
\input{10_conclusion}

%


{
\balance
\bibliographystyle{abbrv}
\bibliography{reference}
}

\clearpage
\newpage
\input{appendix}

\end{document}

%% file: 1_introduction.tex
\section{Introduction}

HTTPS provides the foundation for secure web communication, however, configuring
HTTPS securely has proven to be notoriously difficult~\cite{kotzias2018}. This
is largely due to the complexity of the underlying TLS transport protocol, which
requires server operators to make a wide range of configuration decisions. Most
web servers do not ship with optimal TLS settings, and many come with insecure
options configured by default. To make matters worse, there are hundreds of
options to select from. As of early 2020, there are over 360~cryptographic
ciphers, 44~elliptic curves, 40~protocol extensions, and 6~protocol
versions~\cite{tlsparams}. Seemingly benign settings like session ticket
lifetime have nuanced security implications~\cite{springall2016measuring,
adrian2015}, and recommendations continually change as vulnerabilities
(e.g.,~\cite{freakblog, moller2014, drown2016, adrian2015, cve2009, adrian2015})
are discovered.

The security community has released hundreds of configuration
guides~\cite{mozillaSSL, appliedcrypto}, online tools for verifying
configuration correctness~\cite{ssllabs}, and third-party software to
automatically configure servers~\cite{certbot}. Despite these solutions, large swaths of
the web have been vulnerable to attacks because they continue to use insecure
configurations. For example, when the FREAK attack was disclosed in 2015, 10\%
of popular websites and 37\% of browser-trusted websites still supported
export-grade ciphers, which had been deprecated decades
earlier~\cite{freakblog}. In this paper, we try to better understand whether
servers remain misconfigured today, and if so, why.

We start by empirically evaluating the security posture of the Tranco Top
Million sites~\cite{pochat2018tranco}. We reverse-engineer the configuration of
each website by performing a series of TLS handshakes with varying cryptographic
parameters and grade the user-configurable components of each configuration.
We find that approximately 40\% of sites have optimal settings and 90\% of sites have secure
settings. While this initially appears promising, these aggregate statistics mask 
the coexistence of two
distinct ecosystems with dramatically different security postures. One ecosystem
consists of optimal configurations defined by SaaS (Software as a Service)
providers that use secure-by-default options for all hosted sites. The other
consists of individual operators who are likely manually configuring servers and
struggling to do so securely. Unfortunately, the adoption of SaaS masks that
operators still cannot achieve secure configurations, despite aggregate
ecosystem health improving.\looseness=-1

To better understand why individually configured sites are much less likely to be
secure, we quantify the influence of several factors on configuration security
that operators make decisions about when manually setting up TLS: provider, 
web server software, and online recommendations. We investigate the default
security of the two most popular web servers, Apache and Nginx, as 
well as online recommendations for configuring them on Ubuntu. 
We find that most online recommendations are incomplete and
insecure: 89\% recommend TLS~1.0, 55\% deprecated ciphers, 28\% insecure
ciphers, and 8\% are vulnerable to known attacks.  Similarly, defaults provided by Nginx
and Apache provide sub-optimal security. Our work suggests that the current
approach of sub-optimal defaults along with online recommendations that guide 
operators to fix those settings, is fundamentally broken. We echo previous studies' calls for
fixing systems to be secure by default~\cite{krombholz2017,bernhard2019use}. We
also encourage future studies to consider SaaS services separately from
individually-configured services since aggregate statistics can otherwise be
misleading.

%% file: 9_related.tex
\section{Related Work}
\label{sec:related}

The HTTPS ecosystem and Web PKI has been subject to much attention and there is
a large body of prior work analyzing server usability and
misconfiguration~\cite{bernhard2019use, krombholz2017}, TLS clients and
interception~\cite{durumeric2017security, husak2015, georgiev2012, decarne2016},
server configuration~\cite{kotzias2018, durumeric2015search, durumeric2015,
springall2016measuring, mayer2016}, TLS attacks~\cite{valenta2018search,
bhargavan2016, drown2016, adrian2015,beurdouche2015, valenta2015, garman2015,
moller2014, durumeric2014heart, fardan2013, fardan2013rc4, crime2012}, and
Web PKI~\cite{durumeric2013analysis, amann2013, clark2013, holz2011ssl,
vandersloot2016}. We highlight relevant prior work below.

\paragraph{HTTPS Deployment Difficulty} In 2014, Fahl et~al.\ investigated
common reasons that websites have certificate errors~\cite{fahl2014eve}. Later,
in 2017, Krombholz et~al.~\cite{krombholz2017} showed that technically
proficient users encounter challenges deploying TLS correctly. In 2019, Bernhard
et~al.~\cite{bernhard2019use} analyzed how 10~operators deployed HTTPS on
Apache~2 with and without Let's Encrypt~\cite{aas2019let}. Both Krombholz and
Bernhard use Qualys SSL Labs server test~\cite{ssllabs} to grade deployments.
Bernhard did not find significant evidence to suggest that Let's Encrypt led to
more secure HTTPS deployments. Our study is complimentary and focuses on
real-world deployment difficulties rather than analyzing laboratory
participants.

\paragraph{TLS Measurements} Felt et~al. \ \cite{feltHTTPSadoption} track
the evolution of HTTPS adoption on the Web from 2016-2017 from both a user 
perspective (via aggregate user metrics from Google Chrome and Mozilla Firefox), 
as well as server support for HTTPS among top and long-tail websites.
Kotzias et~al.\ \cite{kotzias2018} fingerprint TLS clients connecting  
to servers at several academic institutions 
in North America from 2012-2017 and document trends in support for cipher suites,
protocol versions, and vulnerabilities over time. They similarly find a 
long tail in TLS deployment that suggests the demand for backwards compatibility. 
Their results provide insights into popular client configurations incoming 
into servers in a certain geographical area, such as browsers and OS-provided 
libraries, but provides limited visibility into the server ecosystem, which is the 
focus of our study. More recently, Holz et~al.\ \cite{holz2020tracking} track 
the deployment of TLS 1.3 from 2017-2019--a small subset of our measurements.
ICSI SSL Notary~\cite{amann2012extracting} similarly provided high level 
statistics of  TLS sessions from 10 participating organizations.
SSL Pulse~\cite{sslpulse} is a dashboard for monitoring the quality
of TLS support over time across 150K HTTPS websites sample from the Alexa Top
Million. For this subset of sites, they present the distribution of grades
as well as support for a number of individual directives (e.g.,  protocol support, 
ciphers, key exchange strength, etc.) and the number of sites being vulnerable 
to various attacks (e.g., DROWN, ROBOT, POODLE, etc.)
Most recently, Lee et al. investigate the spatial differences in TLS configurations and security of 7M domains hosted on content delivery networks (CDNs) and other web hosting services. They find that it is possible to redirect TLS handshake messages to weak TLS servers of which both the origin server and the client may not be aware of~\cite{joonheeSpatial}.

While measuring the support for directives is helpful in assisting operators 
in understanding the deprecation of specific features, they do not provide insight 
into the higher level decisions made by operators (e.g., to keep system defaults, 
to follow recommendations, or to migrate to a cloud provider). Rather than focus 
on individual components Qualys' SSL Labs Test and SSL Pulse~\cite{sslpulse} do, we consider a site's configuration as our unit of analysis.
Our focus is on the various sources of influence on the security of the HTTPS ecosystem 
(defaults, providers, and online recommendations).  While we analyze specific 
configuration choices in order to determine whether a server is set up securely, 
this is a first step in understanding how operators arrive at a configuration, 
rather than the end result of our analysis. As far as we are aware, no prior
work has analyzed server operator decisions at scale or with this intent. 
As we will show, analyzing 
individual directives provides a misleading picture of the security posture 
of the HTTPS ecosystem, as it masks many notable differences among web 
server software and provider.

\paragraph{SSL Grading} Qualys' SSL Server Test~\cite{ssllabs} provides numeric
and letter grades for SSL servers. Qualys' scheme~\cite{ssllabsgrading1,
ssllabsgrading2} grades sites on their protocol support, cipher suite support
and strength, key exchange support and strength, and TLS vulnerabilities. Our
grading scheme is based on SSL Lab's, but differs in a few ways. First, Qualys
rates servers on their certificate and HTTP configuration in addition to TLS
configuration. We do not grade servers on these categories as our focus is on
TLS directives that are user-configurable for servers like Nginx and Apache.
Certificate values are not configured by server operators, and the introduction
of the CA/Browser Forum Baseline Requirements has resulted in certificates being
consistently issued by authorities~\cite{kumar2018tracking}.  Our grading
schemes are comparable in leniency: Qualys' A+ and A grades roughly correspond
with our A grade, their A and B grades with our Bs, their Cs and Ds with our Cs,
and their Fs with  our Fs.

%% file: 2_background.tex
\section{Background}

\paragraph{TLS Handshake} In HTTPS, communication is encrypted using the
Transport Layer Security (TLS) protocol~\cite{rfc5246}. TLS handshakes are
initiated when the client sends a \texttt{ClientHello} message, which specifies
the TLS version, cipher suites, and extensions that the client supports. Each
cipher suite consists of four components: authentication algorithm, key exchange
method, symmetric cipher, and message authentication code (MAC). The client may
choose to list these cipher suites in its preferred order. The server replies to
the client with a \texttt{ServerHello} message that contains the server's
selection of TLS options (e.g., protocol version, cipher suite, extensions) from
the client's presented list of support. For a more detailed reference, we refer
the readers to RFC~5246.

\paragraph{Web Server Configuration} Server operators are faced with several
decisions when configuring a web server.  While some operators choose to host
websites on cloud providers like Cloudflare, which leave a limited amount of
configuration options for users, others use popular open source software such as
Apache~\cite{apache}, Nginx~\cite{nginx}, and Lighttpd~\cite{lighttpd}, which
come equipped with a default configuration. Beyond acquiring and installing a
certificate, operators often choose to manually configure the default settings,
by enabling more secure features such as server cipher preference and disabling
insecure ones like RC4 ciphers. Operators can draw on several sources for
guidance in this process: official server specific
documentation~\cite{nginxdocs, apachedocs}, online recommendations such as the
Mozilla SSL Configuration Generator~\cite{mozillaSSL}, personal blogs by
security experts, and automatic tools that configure servers~\cite{certbot}.
Online tests such as SSL Labs' SSL Server Test~\cite{ssllabs} provide a means of
testing the security of TLS configurations and identifying security issues.

%% file: 3_methodology.tex
\section{Methodology and Data Collection}

We start our analysis by collecting and grading the HTTPS configurations of the
Tranco Top Million websites. In this section, we describe our HTTPS scanner and
grading methodology. To note, we build a scanner instead of using SSL Labs' 
due to the scale of our study---we scan several million sites, 
which would not have been possible in a reasonable time frame using SSL Labs' API. 

\subsection{Defining a Configuration}

We define a configuration in terms of the TLS options that are configurable in
the four most common user-configurable web sites: Nginx, Apache, Lighttpd, and
IIS. This includes TLS version, cipher suites, session ID resumption, extension
session ticket, ticket lifetime hint, compression support, server cipher
preference, Diffie-Hellman (DH) group size, and Diffie-Hellman group, but
excludes library-controlled settings like TLS Heartbeat support. To avoid the
exponential explosion of variable space and because cipher suites are typically
introduced by cipher, we analyze cipher components (e.g., ECDHE, 3DES, RC4, MD5)
rather than individual TLS cipher suites. All features are binary indicators,
with the exception of DH group size and ticket lifetime hint, which we
discretize.

\subsection{Collecting Server Configurations}

To collect configurations, we introduce a scanner that reverse engineers the
HTTPS settings of websites by performing a series of HTTPS handshakes that
present varying sets of cryptographic parameters. We detail the scanner and
handshakes below:

\paragraph{HTTP(S) Support} We first attempt a TLS handshake that emulates a
modern browser by offering TLS~1.2 support and the union of ciphers present in
the recent versions of Chrome, Safari, Firefox, and Edge.\footnote{We
specifically use the TLS~1.2-compatible cipher suites specified by Chrome~65,
Safari~13.0.1, Firefox~66, and Microsoft Edge~18.18363.} If this handshake
succeeds, we attempt an HTTP \texttt{GET\,/} over TLS\@. We exclude hosts that
do not complete these two handshakes from our study as they would not be
accessible to normal users.
We also note the certificate signature algorithm to decide whether to offer
ECDSA or RSA ciphers in future handshakes.

\paragraph{TLS Version} We check SSLv3--TLSv1.2 protocol support by iteratively
sending TLS \emph{Client Hello} messages that indicate support for progressively
older TLS versions. For every message, we present the set of versions older than
the server's last indicated support. For example, if a server selects TLSv1.1 in
response to our TLS1.2~handshake, we would present TLSv1, and if successful,
SSLv3. We offer commonly supported cipher suites and curves in each handshake.
We test support for SSLv2 and TLSv1.3 separately because they use different
handshakes or mechanisms for indicating version support.

\paragraph{Cipher Suite Support} We test support for DES, 3DES, RC4, IDEA, SEED,
Camellia, ARIA, ChaCha, and AES symmetric ciphers, MD5, SHA-1, SHA-256, and
SHA-384 MACs, and NULL or EXPORT components.  We do not test for widespread
support of PSK, SRP, ECCPWD, KRB, DSS, DH, or ECDH-based key exchange methods,
MD2 MACs, or Anon signature algorithms, because we find that less than 0.1\% of
sites in the Tranco Top Million support these algorithms.\footnote{Prior to
doing our large-scale scan of Tranco and CT, we scanned Tranco for support of
the comprehensive set of ciphers supported by IANA.} To measure cipher suite
support, we send a Client Hello that indicates support for all cipher suites
compatible with the presented certificate (i.e., ECDSA or RSA) and iteratively
remove the server selected cipher until the handshake fails.  Although browsers
typically present fewer than 30~cipher suites, we find that including 97~cipher
suites (the maximum we offer in our
first handshake) does not preclude any server from completing a handshake. 

\paragraph{Extensions} We test support for TLS extensions by sending a single
handshake that indicates support for the following extensions: server name,
heartbeat, session ticket, ALPN, OCSP stapling, secure renegotiation, extended
master secret, and signed certificate timestamp, and checking for selection in
the \texttt{Server Hello}.  If the server indicates support for heartbeat, we
also test for the Heartbleed vulnerability.

\paragraph{TLS Compression} We attempt a handshake with TLS compression methods
\texttt{deflate} and \texttt{LZS}~\cite{compression}.

\paragraph{Session Resumption} We test support for session ID resumption and
session tickets by sending two handshakes each, to establish and attempt to
resume sessions.

\smallskip \noindent We implemented our scanner in Go, using the ZCrypto TLS
library~\cite{zcrypto}. We have released our scanner at \url{https://github.com/stanford-esrg/tls-webserver-configuration-scanner} under the Apache 2.0 license. In total, our algorithm completes
14--93 handshakes per server. The large range is primarily due to the large
number of cipher suites that a server can theoretically support. On average, we
perform 40--43~handshakes per site.

\subsection{Grading Configurations}

Inspired by Qualys' SSL Server Test~\cite{ssllabs}, we define a grading scheme
to quantitatively compare the security provided by different configurations (see
Section~\ref{sec:related} for a comparison between our  algorithm with Qualys'
SSL Server Test).  We assign a grade of A (optimal), B (non-optimal but safe), C
(weak, vulnerable to academic attacks), and F (severely broken, simple MITM
possible) grades to websites. We grade sites based on 7~categories and assign
the minimum score across the categories:

\paragraph{Protocol Support} To receive an A, sites must support TLSv1.2 or 1.3
and must not support SSLv2 or 3. Sites that support SSLv2 receive an F; sites
that do not support TLSv1.2 or~1.3 receive a C\@.

\paragraph{Key Exchange} Sites must only support ECDHE key exchange for an A;
sites must support a minimum 2048-bit DHE group or an uncommon 1024-bit group
for a B\@. Sites with 768-bit DHE groups receive a C, and sites with smaller DHE
groups receive an F\@.

\paragraph{Symmetric Ciphers \& MAC Algorithm} To receive an A, sites must
support AEAD ciphers, and not support any of: Camellia, ARIA, IDEA, SEED, RC4,
or DES ciphers, MD5, NULL or EXPORT ciphers. For a B, sites must not support any
of: RC4 or DES ciphers, MD5 MACs, NULL or EXPORT components. To receive a C,
sites must not support DES ciphers or any NULL or EXPORT components. Otherwise,
sites will receive an F for supporting DES ciphers or any NULL or EXPORT
components.

\paragraph{Preferred Cipher} Sites must support server cipher preference
and must prefer an AEAD cipher and Perfect Forward Secrecy (PFS)
to receive an A\@. All other sites receive a B\@.

\paragraph{Compression} Sites that support TLS compression receive a C, otherwise,
they receive an A\@.

\paragraph{Ticket Lifetimes} Sites with session ticket hints longer than a week
receive a C; 1--7 days, B; and under 24 hours, A\@. While a session ticket hint
can differ from the ticket lifetime, Springall et~al.\ showed that they are
nearly always the same in practice~\cite{springall2016measuring}.

\paragraph{Vulnerabilities} Sites vulnerable to CRIME\cite{crime2012},
POODLE\cite{moller2014}, or FREAK\cite{valenta2015} receive a C\@. Sites
vulnerable to Heartbleed receive an F\@. Sites not vulnerable to any TLS attacks
to receive an A\@.\footnote{ We do not consider elliptic curve support in our
definition of a configuration as there are no known curve-based vulnerabilities
(despite downgrade attacks~\cite{curveswap}) and there have been bugs in Nginx
related to curve selection, where even if specific curves are specified, they
may not end up being used~\cite{mozilla76} \cite{mozilla189}.}

\subsection{Data Collection and Annotation}

We analyze 1.16M~websites that appeared in the Tranco Top
Million~\cite{pochat2018tranco} in August 2019. We performed our scan between
September 4--8, 2019 from 3~servers at Stanford University. We completed handshakes over
a 102~hour period and waited a random 0--2400~second period between each
handshake in order to prevent overloading sites. Of the 1.16M~candidate sites,
we were able to resolve 1.06M names and successfully complete a TCP handshake on
port 443 with 890K\@. Of those, 678K sites presented a browser-trusted
certificate that matched the name on the website and responded to an HTTPS
\texttt{GET\,/} request.
In the event of a non-TLS related error, our scanner attempts another handshake
before proceeding. This extra attempt accounts for potential network loss. We
followed the best practices set forth by Durumeric et
al.~\cite{durumeric2013zmap}, configuring the HTTP page on scan hosts to
redirect to a website that explained our study. We received only one exclusion
request.

We extract the server software from the HTTP \texttt{Server}
header~\cite{rfc2616} and use MaxMind GeoIP2 GeoLite~\cite{geolite} to geolocate 
sites at the country-level.  We identify Content Delivery Networks (CDNs) and
Software as a Service (SaaS) providers from their Autonomous System (AS) number
and name.

Finally, to understand whether different types of sites have differing security
profiles, we map a random sample of 100K~sites to the Alexa Top
Sites~\cite{alexa} 17~categories.\footnote{We limit our analysis to 100K~sites
instead of the complete list due to resource constraints. Alexa Top Sites
categorizes websites into the following categories: Adult, Arts, Business,
Computers, Games, Health, Home, Kids and Teens, News, Recreation, Reference,
Regional, Science, Shopping, Society, Sports, and World.} We were able to
successfully map approximately 67\% of our sample to categories.

%% file: 4_ecosystem.tex
\section{HTTPS Ecosystem Security}
\label{sec:ecosystem}

\begin{figure}[t]
	\includegraphics[width=0.5\textwidth]{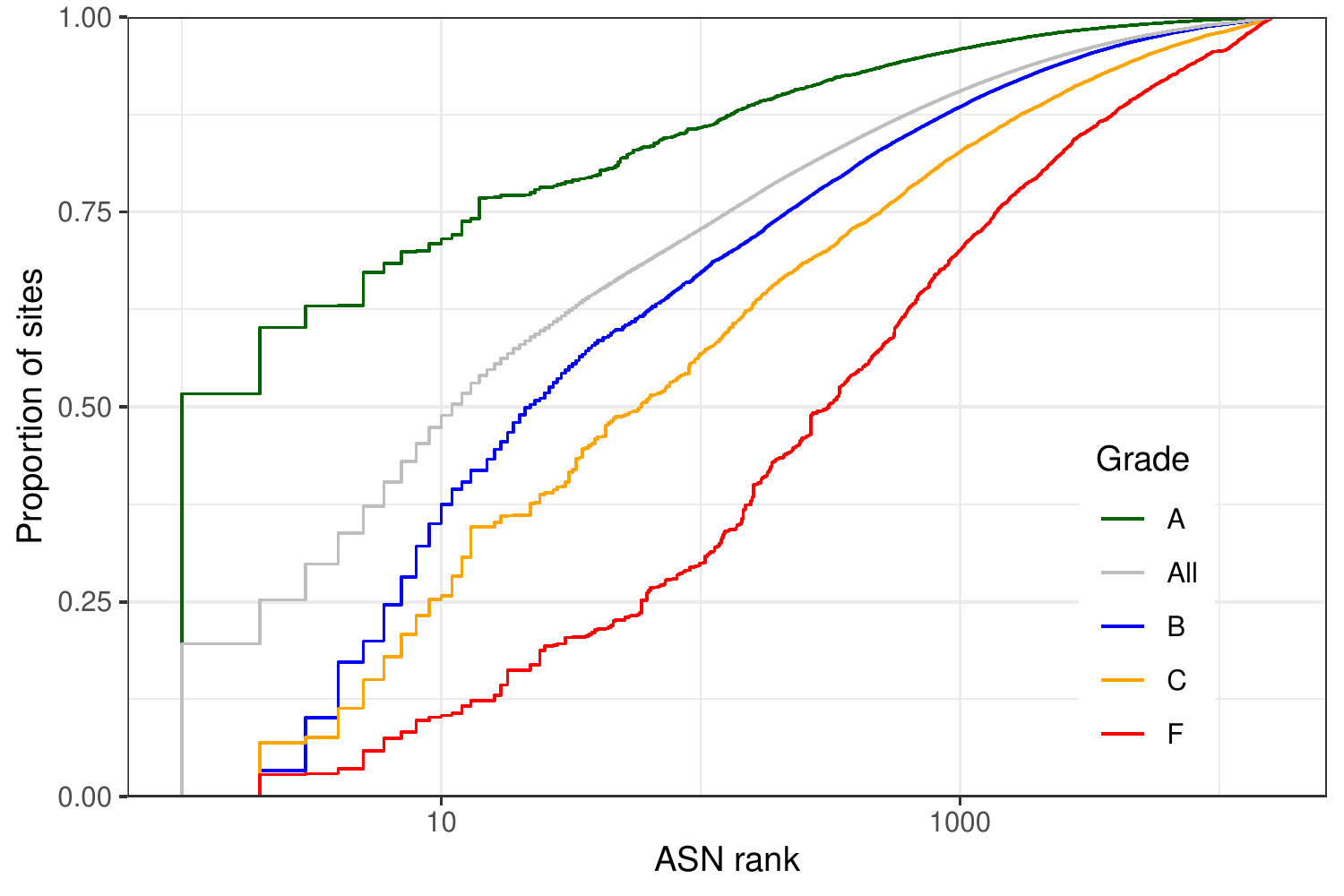}
	\caption{\textbf{CDF of site count by grade and ASN rank.}---Cloudflare's dominance and its small set of A-grade configuration is responsible for approximately 50\% of As, lifting the average grade for all sites.
}
\label{fig:asn_cdf_tranco}
\end{figure}

Across the 678K~sites in our study, we find 7.7K~unique configurations in over
16K~ASes. The most popular configuration accounts for 14\% of sites and belongs
to Cloudflare. The long tail begins thereafter, with the next most popular
configuration being adopted by only 3.3\% of sites and hosted across 1,421~ASes.
Nearly 90\% of all sites have secure HTTPS settings (A or B).  Nearly 40\%
receive an optimal A grade, 10\% receive Cs, and only 1\% Fs. While this
initially appears very promising, the aggregate number masks notable differences
in the distribution of grades across server software and provider, which we
detail in the remainder of this section.

\subsection{CDN Providers}

A small number of CDNs and unique configurations account for the majority of
secure sites (A or B grade).  Ten ASes account for approximately 72\% of sites
with As and 38\% of Bs; 100~ASes account for 86\% of As and 67\% of Bs. At first
glance, this concentration may not appear surprising. Infrastructure providers
host a large number of \emph{all} sites.  The top~10 and vast majority of the
top~100 ASes belong to SaaS companies, cloud providers, and CDNs. Indeed nearly
73\% of all websites---regardless of grade---are hosted in 100~ASes and 33\% of
sites are hosted in 10~ASes. However, as can be seen in
Figure~\ref{fig:asn_cdf_tranco}, sites with poor security do not see this
concentration, and are widely spread across a large number of ASes: 100~ASes
account for only 57\% of Cs and 30\% of Fs.

\begin{figure}[t]
	\includegraphics[width=0.5\textwidth]{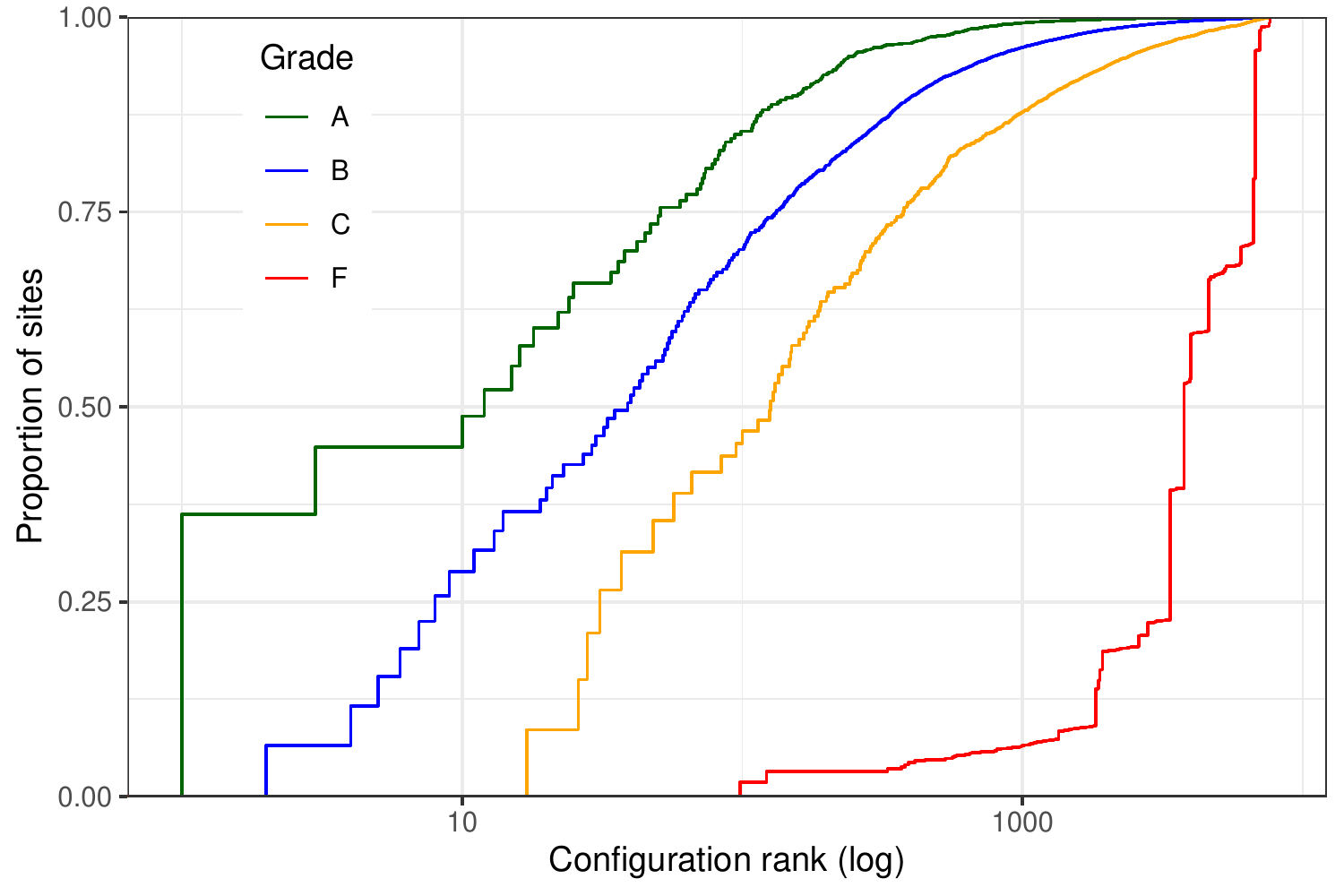}
	\caption{\textbf{CDF of site count by grade and configuration rank.}---
	Sites receiving As concentrate in higher-ranking configurations, whereas
	configurations that are less commonly used are more insecure.
}
\label{fig:configuration_cdf_tranco}
\end{figure}

We investigate the sites in the top 100~ASes and find that in many cases, there
is a single or small handful of configurations in popular SaaS/CDN/cloud
networks that account for a significant fraction of sites. This suggests that
providers exert influence on operators' choice of configuration options, whether
it be via a default configuration, limiting configuration options, or providing
clear recommendations that operators use in practice. Within the top ten ASes,
the most popular configuration in each AS is used by a significant proportion of
sites, and in each case, it is secure (i.e., receives either an A or B grade).
At the extreme, a single Cloudflare configuration accounts for 52\% of
\emph{all} sites in our study that receive an A grade. The top
100~configurations that receive an A grade account for 97\% of all A websites.
In contrast, the top 100~configurations that receive an F grade account for only
18\% of sites with Fs.

In some cases, this is due to the provider offering their own SaaS service, even
when the top configuration does not account for the majority of sites on the
provider. For example, the most popular configuration in Amazon ASes belongs to
Cloudfront, which offers clients six possible security policies to choose
from~\cite{amazon-policies}. The dominant configuration accounts for 14\% of
Amazon sites and 5.6\% of all sites in Tranco. We also see evidence of smaller
SaaS providers operating within the cloud providers. For example, Google's most
dominant configuration is used in 39\% of its sites and belongs to a hosted
third-party provider, WP Engine~\cite{wpengine,wp-google}.  We empirically infer
the default configurations for providers, and provide the five most popular
configurations within each of the top 10~ASes in Tranco in
Table~\ref{tab:saas_defaults}.  The default TLS settings are typically not
advertised online and we found it intractable to create accounts for each
provider, especially as each provider typically offers multiple plans, including
enterprise policies. Some CDNs may provide more or less than five default
configurations, however, as can be seen in Table~\ref{tab:saas_defaults}, any
defaults we have missed are only used by a small proportion of sites in the tail
of the distribution and are thus less impactful on the ecosystem as a whole.  We
find that all of the dominant configurations are secure, receiving either an A
or B grade.

While we cannot exactly quantify the improvement in security that SaaS providers
are responsible for due to the tremendous number of providers globally, we
observe an unambiguous positive contribution to improving the security of the
overall HTTPS ecosystem from even the ten most popular. There is a clear disparity
in the security profile for sites hosted on large, well-known SaaS providers
versus individually-configured sites, which are likely manually configured
servers by operators.  We confirm this finding more generally by considering the
top configurations independent of the AS they are in.  A pairwise T-test with a
Bonferroni adjustment for multiple comparisons confirms that configuration
dominance (i.e., the proportion of sites that use a configuration) is correlated
with grade: the unique configurations that receive higher grades are used by a
larger number of sites than configurations that are insecure.
Figure~\ref{fig:configuration_cdf_tranco} shows the complete distributions of
configuration rank in terms of the proportion of sites using each configuration.
At the other extreme, we find that for uniquely configured sites, the odds of
the configuration being insecure (receiving a grade of C or F) is 1.30~times
higher than being secure---57\% versus 43\% of uniquely configured sites are
insecure.  The same qualitative result holds even if we relax this threshold and
consider configurations used in more than one site, although the effect is
attenuated.  For example if considering configurations used in at least
10~sites, the odds of having an insecure configuration is 1.09 times that of
having a secure configuration.

\begin{table}[t]
\footnotesize
\setlength{\tabcolsep}{4pt}
\centering
\input{tables/grade_drop.tex}

\label{tab:grade_drop_reasons}
\end{table}

\subsection{Drivers of Insecurity}

Next, we investigate the main reasons why sites are insecure, which are 
summarized in Table~\ref{tab:grade_drop_reasons}. Of sites that
received a B, the primary reasons why servers were downgraded was a lack of
cipher preference (12\% of sites) and support for at least one suboptimal cipher
(13\%). Sites were downgraded to a C for key exchange (20\%  of sites used 1024
bit common DH parameters and are susceptible to Logjam) and support for ciphers
(40\% of sites supported RC4 and 32\% supported MD5 MACs). To prevent a majority
of downgrades to C, servers can generate fresh 2048 bit DH parameters with
OpenSSL or use Mozilla's predefined \texttt{ffdhe2048}
parameter~\cite{mozillaSSL} and remove support for RC4 and MD5 macs. Last, of
sites that were downgraded to an F, 44\% were downgraded due to support for
SSLv2, 29\% were downgraded due to a weak key exchange strength, 29\% were
downgraded due to supporting insecure ciphers (80--89\% of these supported
export grade ciphers, NULL cipher components, and DES ciphers), and 5\% were due
to having vulnerabilities (all sites receiving an F were vulnerable to
Heartbleed, 90\% supported compression and CBC ciphers).  Patching Heartbleed
requires an update of the cryptographic library.  A comprehensive description of
individual directive support is included in the Appendix.  We caution that our
analysis only captures a snapshot of the current state of the HTTPS ecosystem,
and results may be volatile over time.

Of sites that were downgraded to B because of their cipher preference, 74\% did
not have server cipher preference and 56\% first negotiated with non-AEAD
ciphers. Of servers that were downgraded to B because of their cipher suite
support, over 74--99\% supported older, uncommon, and non-AEAD ciphers: ARIA,
Camellia, SEED, or IDEA\@. This suggests that enabling server cipher preference
and removing support for uncommon ciphers are simple changes that could upgrade
servers to optimal security without affecting client compatibility. While we
cannot confirm for certain whether an insecure site is left so intentionally,
for example, in order to support legacy clients, previous work has documented
the difficulties that operators face when configuring servers~\cite{fahl2014eve,
krombholz2017, bernhard2019use}, so it is likely that a significant portion, if
not all, of these issues are attributable to configuration mistakes or are
configurations that have become outdated over time. Further, many of the
misconfigurations made do not improve compatibility with any version of
mainstream browsers.

%% file: tables/grade_drop.tex
\begin{tabular}{r%
	l@{\hskip -4pt}%
	l@{\hskip -4pt}%
	l@{\hskip -4pt}%
	l@{\hskip -4pt}%
	l@{\hskip -4pt}%
	l@{\hskip -4pt}%
	l@{\hskip -4pt}%
	l@{\hskip -4pt}%
	rlllllllll}
\toprule
& \multicolumn{2}{c}{$\rightarrow$ B } &   \multicolumn{2}{c}{$\rightarrow$ C }   &  \multicolumn{4}{c}{$\rightarrow$ F } \\
\cmidrule(r){2-3} \cmidrule(r){4-5} \cmidrule(r){6-10}
&
\begin{turn}{65} Preferred $\;\;$   \end{turn} &
\begin{turn}{65} Ciphers $\;\;\;\;\;\;\;\;$   \end{turn} &
\begin{turn}{65} KEX $\;\;\;\;\;\;\;\;$ 	\end{turn} &
\begin{turn}{65} Ciphers  $\;\;\;\;\;\;\;\;$\end{turn} &
\begin{turn}{65} Protocol $\;\;\;\;$ 	\end{turn} &
\begin{turn}{65} KEX $\;\;\;\;\;\;\;\;$ 	\end{turn} &
\begin{turn}{65} Ciphers  \end{turn} &
\begin{turn}{65} Vulns  \end{turn} \\
\midrule
  Tranco  & 12 & 13 & 20 & 40 & 44 & 0  & 29 & 5 & \\
\bottomrule
\end{tabular}
\caption{\textbf{Primary reasons for grade downgrades}---Primary factors contributing to downgrades from each grade to the next lower one. Each value represents the proportion of sites downgraded to each grade (from better grades) by reason.
}

%% file: 5_model.tex
\subsection{Ordinal model
\label{ordinalmodel}}

To more quantitatively confirm our finding that cloud providers are primarily
responsible for secure configurations, we model the relationship between server
security and several variables related to the provider, server software,
geographic location, site content, and site popularity in Tranco.  For example,
it could be that popular sites are more secure than unpopular ones, and popular
sites use SaaS providers.
We exclude Cloudflare sites as there are few configuration choices left to the
user and little variance in the grade---virtually all sites obtain an A.

\begin{table}[t]
\setlength{\tabcolsep}{5pt}
\small
\centering
\begin{subfigure}[t]{0.58\columnwidth}
\centering
\begin{tabular}{lrrrr}
\toprule
  				& F/C &  B & A \\ 
\midrule
Nginx  	 	& 	0.94 &  0.99 & 1.07  \\
IIS    		&   1.10 &  1.01 & 0.90 \\
Litespeed   &   0.73 &  0.81 & 1.71 \\
OpenResty   &   0.88 &  0.95 & 1.20 \\
CPanel    	&   1.10 &  1.00 & 0.90 \\
BigIP    	&   0.87 &  0.95 & 1.20 \\
Cloudfront  &   0.72 &  0.79 & 1.77 \\
Varnish    	&   0.83 &  0.92 & 1.32 \\
ATS    		&   0.92 &  0.98 & 1.11 \\
AWSelb/2.0  &   0.73 &  0.79 & 1.72 \\
SquareSpace &   0.98 &  1.00 & 1.03 \\
Akamai   	&   0.75 &  0.82 & 1.64 \\
GHS			&   0.97 & 0.99 & 1.03 \\
Other     	&   0.87 & 0.96 & 1.20 \\
\bottomrule
\end{tabular}
\end{subfigure}
\hfill
\begin{subfigure}[t]{0.4\columnwidth}
\centering
\begin{tabular}{lrrrr}
\toprule
  				& F/C &  B & A \\ 
\midrule
.org        &  0.99 & 1.00 & 1.01   \\
.net        &  1.01 & 1.00 & 0.99   \\
.de         &  0.98 & 1.00 & 1.03   \\
.uk         &  0.99 & 1.00 & 1.02   \\
.jp         &  1.05 & 1.01 & 0.94   \\
.ae         &  1.02 & 1.00 & 0.98   \\
.au       	&  0.97 & 0.99 & 1.03  \\
.ir 		&  1.03 & 1.00 & 0.97   \\
.it 		&  1.03	& 1.00 & 0.97   \\
.co 		&  0.97	& 0.99 & 1.04   \\
.ua 		&  0.96	& 0.99 & 1.05   \\
.io 		&  0.94	& 0.98 & 1.09  \\
.es 		&  1.02	& 1.00 & 0.98  \\
.vn 		&  1.05	& 1.01 & 0.95  \\
\bottomrule
\end{tabular}
\end{subfigure}
\caption{
Marginal effect sizes in terms of odds for ordinal logistic regression.
Only significant predictors, with an absolute effect size greater than 1\% are shown.
The baseline categories for server is Apache, and ``.com''.
All coefficient are significant at the 0.05 confidence level.
}
\label{tab:ordinal_logistic_regression}
\end{table}

We treat the set of sites measured as a random sample of the most popular sites.
We fit a generalized ordered logistic regression model using \texttt{R}'s
\texttt{oglmx} package~\cite{oglmx} to understand the relationship between a
server's grade and several variables related to the server software, geographic
location, site content, and site popularity in Tranco. As is usual in these
contexts, we assume that, underlying the observed discrete grade categories of a
site i, $Y_i$, there is a continuous latent variable $\tilde{Y_i}$ that captures
the security posture of a site.  We assume that the mean of the underlying
latent variable may be modeled as a linear combination of explanatory variables:
\begin{equation}
\tilde{Y_i}  = \beta_0 + \beta_{server} \mbox{s}_i + \beta_{tld} \mbox{t}_i +  \beta_{median} \mbox{m}_i + \beta_{sd} \mbox{d}_i + \epsilon_i
\end{equation}

\noindent where $\beta_0$ is an intercept term, $s_i$ represents the server
software, $t_i$ represents the top level domain of the site; $m_i$ and $d_i$
represent the median and standard deviation of the site rank across the one
month period, respectively; and $\epsilon_i$ is a mean zero error assumed to be
logistically distributed across observations.

We find that server software and TLDs (especially country-level TLDs) are highly
correlated with grade, while the other covariates have minimal effect.
Microsoft-IIS and CPanel in particular stand out. The odds of obtaining a high
grade decrease, from 1.10~times greater for a F/C, to 0.88~lower for an A. There
is some evidence that top level domain and geographical location of the site
matters. Sites with Japanese and Vietnamese TLDs are more likely to obtain a low
grade.  The odds of obtaining an F/C is 1.05 times higher for Japan and Vietnam,
and the odds of obtaining an A is 0.94 lower for Japan, and 0.95 lower for
Vietnam. In contrast, sites with .io, .co, and .ua are more likely to get an A,
with odds of 1.09, 1.05, and 1.04 respectively.  The popularity of the site does
not seem to matter. The median rank was not found to be significant, and the
effect sizes for standard deviation of rank were small, even when significant.
In contrast, sites hosted on LiteSpeed, BigIP, Varnish, as well as several cloud
providers: Amazon CloudFront, AWS elb/2.0, and Akamai increase the odds of
obtaining an A.  The effects are large and significant.  For example, the odds
of obtaining an A (as opposed to a lower grade) is 1.77 times higher for a site
hosted on Cloudfront, than for one one hosted on Apache. The complete model
output including significance results are given in
Table~\ref{tab:complete_model_results}.

%% file: 6_defaults.tex
\subsection{Ubuntu Defaults\label{defaults}}

To better understand why so many individually configured sites receive B, C, and
F grades, we investigate the default settings and online recommendation guides
for the two most popular web servers, Apache and Nginx. Together, Apache and
Nginx account for 70\% of non-Cloudflare Tranco sites.  We primarily investigate
Ubuntu defaults as Ubuntu is the most commonly identifiable OS, accounting for
48\% of sites that identify an OS in their server string in Tranco.~\footnote{
We are able to identify the OS from the server string on only 10\% of sites in
Tranco}. To identify default Ubuntu settings, we installed the last four Ubuntu
LTS releases (12.04--18.04) with all compatible combinations of packaged OpenSSL
(1.0.1--1.1.1) and Nginx (1.1.19--1.14.0) or Apache (2.2.22--2.4.29) packages.
In total, we analyze 2,069~combinations and extract 11~unique configurations
(Table~\ref{tab:defaults}) consistent with our definition of a configuration.

\begin{table*}
\small
\setlength{\tabcolsep}{4pt}
\centering
\input{tables/defaults.tex}

\caption{
Default configurations for Ubuntu and OpenSSL on Nginx and Apache.
A circle represents support, an empty circle indicates no support.
Values represent the proportion of domains supporting each default out of the subset of sites running Ubuntu and each server in Tranco and CT (n=15K Apache and n=9.6K Nginx in Tranco; n=13K Apache and n=9.4K Nginx in CT). Several defaults had multiple server types as indicated by the footnotes:
[1] includes Ubuntu versions 12.04, 14.04;
[2] includes OpenSSL versions 1.1.0g, 1.1.1;
[3] includes OpenSSL versions 1.0.1, 1.0.1f.}
\label{tab:defaults}
\end{table*}

We find that no default receives an A grade; the majority (5/11) receive a
B\@.\footnote{A server supports a TLS feature by default if: (1) the server
software implements the feature, (2) the feature is enabled through compilation,
and (3) the feature is turned on in the configuration file.} All Ubuntu defaults
support TLS 1.0--1.2, session tickets valid for 300~seconds, AES-128, AES-256,
and AES-GCM ciphers, AEAD, CBC, Camellia and ARIA ciphers, RSA key exchange, as
well as SHA-1 MACs. Five defaults on older Ubuntu versions support SSL~3.0 and
two on the most recent version (18.04) support TLS~1.3. Two Apache on Ubuntu
configurations support RC4 and three support 3DES ciphers; no defaults support
DES or IDEA ciphers, MD5 MACs, or NULL or EXPORT components. We note that
default security has improved over time, with the most recent versions obtaining
B-grades. Nevertheless---they still lag behind cloud providers and servers that
offer secure TLS settings by default like Caddy~\cite{caddy} of which 99\% of
sites receive an A grade.

Both Nginx and Apache on Ubuntu fall short of an A because they support uncommon
ciphers (e.g., CAMELLIA). Additionally, Apache configurations are capped at a B
because they do not have server cipher preference. Encouragingly, the latest
versions of Apache on 18.04 have TLSv1.3 support by default.  Nginx on 18.04 has
intermittent support for TLSv1.3. Only a small number of sites in Tranco use
these exact default values---1.7\% of sites.
Of sites that identify as using Nginx, 1.8\% support an Nginx default;
of sites that identify as using Apache, 4\% support an Apache default. It is
unclear why operators change settings that do not improve site security.  It may
be a misguided attempt to improve security, or an attempt to fix unrelated
problems.

The Ubuntu configurations of Nginx and Apache differ in several ways. Nginx on
Ubuntu by default has server cipher preference in 16.04 onward. We note that
this is a Ubuntu specific choice rather than an Nginx default
~\cite{nginxdefaults}, which has server cipher preference turned off.  Other
OSes such as Fedora~\cite{fedoraNginx} and Centos~\cite{centosNginx} do not set
server cipher preference.  Apache on Ubuntu does not use server preference in
its defaults~\cite{apachedefaults}.  Additionally, Apache on Ubuntu by default
supports session ID resumption~\cite{apachedefaults} whereas Nginx on Ubuntu
does not~\cite{nginxdefaults}. Starting with version 18.04, Nginx on Ubuntu also
removed support for finite-field-based Diffie-Hellman key exchange.

%% file: tables/defaults.tex
\begin{tabular*}{\textwidth}{l%
	l@{\hskip -8pt}%
	l@{\hskip -2pt}%
	l@{\hskip -2pt}%
	l@{\hskip -2pt}%
	l@{\hskip -2pt}%
	l@{\hskip -8pt}%
	l@{\hskip -8pt}%
	l@{\hskip -8pt}%
	l@{\hskip -8pt}%
	l@{\hskip -8pt}%
	l@{\hskip -8pt}%
	l@{\hskip -8pt}%
	l@{\hskip -8pt}%
	l@{\hskip -8pt}%
	l@{\hskip -8pt}%
	l@{\hskip -8pt}%
	l@{\hskip -8pt}%
	l@{\hskip -8pt}%
	l@{\hskip -8pt}%
	l@{\hskip -8pt}%
	l@{\hskip -8pt}%
	l@{\hskip -8pt}%
	l@{\hskip -8pt}%
	l@{\hskip -8pt}%
	l@{\hskip -8pt}%
	l@{\hskip -8pt}%
	l@{\hskip -8pt}%
	l@{\hskip -8pt}%
	l@{\hskip -8pt}%
	l@{\hskip -8pt}%
	l@{\hskip -8pt}%
	llllccrrrrrrrrrrrrrrrrrrrrrcrrrrr}
\toprule
\begin{turn}{65} Server \end{turn} &
\begin{turn}{65} Ubuntu ver $\;\;\;\;\;\;\;\;$ \end{turn} &
\begin{turn}{65} OpenSSL $\;\;\;\;\;\;\;\;\;$ \end{turn} &
\begin{turn}{65} Grade $\;\;\;\;\;\;\;\;\;\;$ \end{turn} &
\begin{turn}{65} \% Tranco \end{turn} &
\begin{turn}{65} TSLv1.3 $\;\;\;\;$ \end{turn} &
\begin{turn}{65} TSLv1.2 $\;\;\;\;$ \end{turn}  &
\begin{turn}{65} TSLv1.1 $\;\;\;\;$  \end{turn} &
\begin{turn}{65} TSLv1.0 $\;\;\;\;$  \end{turn} &
\begin{turn}{65} SSLv3.0 $\;\;\;\;$  \end{turn} &
\begin{turn}{65} SSLv2.0 $\;\;\;\;$  \end{turn} &
\begin{turn}{65} Server pref $\;\;\;\;$  \end{turn} &
\begin{turn}{65} Sesssion ID $\;\;\;\;$ \end{turn} &
\begin{turn}{65} Session Ticket   \end{turn} &
\begin{turn}{65} RC4 $\;\;\;\;\;\;\;\;\;\;\;$ \end{turn} &
\begin{turn}{65} DES $\;\;\;\;\;\;\;\;$ \end{turn}&
\begin{turn}{65} 3DES $\;\;\;\;\;\;\;\;$ \end{turn} &
\begin{turn}{65} Export $\;\;\;\;$ \end{turn}&
\begin{turn}{65} ARIA $\;\;\;\;\;\;\;$ \end{turn}&
\begin{turn}{65} Camellia $\;\;\;\;$ \end{turn}&
\begin{turn}{65} IDEA $\;\;\;\;\;\;\;$ \end{turn}&
\begin{turn}{65} SEED $\;\;\;\;$ \end{turn}&
\begin{turn}{65} CHACHA $\;\;\;\;$ \end{turn}&
\begin{turn}{65} AEAD $\;\;\;\;$ \end{turn}&
\begin{turn}{65} AES-GCM $\;\;\;\;$ \end{turn}&
\begin{turn}{65} DH Group $\;\;\;\;\;\;\;\;$ \end{turn}&
\begin{turn}{65} KE:DHE $\;\;\;\;$  \end{turn}&
\begin{turn}{65} KE:ECDHE $\;\;\;\;$ \end{turn}&
\begin{turn}{65} KE:RSA  $\;\;\;\;$ \end{turn}&
\begin{turn}{65} MD5 $\;\;\;\;\;\;\;$  \end{turn}&
\begin{turn}{65} SHA  \end{turn} \\
\midrule
Nginx & 	14.04 & 	1.0.1f	 & C & 	8  &  $\circ$ & $\bullet$ & $\bullet$ & $\bullet$ & $\bullet$ & $\circ$ & $\circ$ & $\circ$ & $\bullet$ & $\circ$ & $\circ$ & $\circ$ & $\circ$ & $\circ$ & $\bullet$ & $\circ$ & $\circ$ & $\circ$ & $\bullet$ & $\bullet$ & 1024	& $\bullet$ & $\bullet$ & $\bullet$ & $\circ$ & $\bullet$ \\
Nginx & 	16.04 & 	1.0.2g	 & C & 	28 	  	& $\circ$ & $\bullet$ & $\bullet$ & $\bullet$ & $\circ$ & $\circ$ & $\bullet$ & $\circ$ & $\bullet$ & $\circ$ & $\circ$ & $\circ$ & $\circ$ & $\circ$ & $\bullet$ & $\circ$ & $\circ$ & $\circ$ & $\bullet$ & $\bullet$ & 1024	& $\bullet$ & $\bullet$ & $\bullet$ & $\circ$ & $\bullet$ \\
Nginx & 	18.04 & 	1.1.0g$^2$	 & B & 	0  	& $\circ$ & $\bullet$ & $\bullet$ & $\bullet$ & $\circ$ & $\circ$ & $\bullet$ & $\circ$ & $\bullet$ & $\circ$ & $\circ$ & $\circ$ & $\circ$ & $\bullet$ & $\bullet$ & $\circ$ & $\circ$ & $\bullet$ & $\bullet$ & $\bullet$ & -	& $\circ$ & $\bullet$ & $\bullet$ & $\circ$ & $\bullet$ \\
Nginx & 	18.04 & 	1.1.0g$^2$	 & B	 & 0   	& $\bullet$ & $\bullet$ & $\bullet$ & $\bullet$ & $\circ$ & $\circ$ & $\bullet$ & $\circ$ & $\bullet$ & $\circ$ & $\circ$ & $\circ$ & $\circ$ & $\bullet$ & $\bullet$ & $\circ$ & $\circ$ & $\bullet$ & $\bullet$ & $\bullet$ &  -	& $\circ$ & $\bullet$ & $\bullet$ & $\circ$ & $\bullet$ \\
\midrule
Apache & 	12.04 & 	1.0.1    & F	 & 2 	& $\circ$ & $\bullet$ & $\bullet$ & $\bullet$ & $\bullet$ & $\circ$ & $\circ$ & $\bullet$ & $\bullet$ & $\bullet$ & $\circ$ & $\bullet$ & $\circ$ & $\circ$ & $\bullet$ & $\circ$ & $\bullet$ & $\circ$ & $\bullet$ & $\bullet$ & 1024	& $\bullet$ & $\circ$ & $\bullet$ & $\circ$ & $\bullet$ \\
Apache & 	12.04$^1$ & 	1.0.1$^3$  &   C/F & 	24 	& $\circ$ & $\bullet$ & $\bullet$ & $\bullet$ & $\bullet$ & $\circ$ & $\circ$ & $\bullet$ & $\bullet$ & $\bullet$ & $\circ$ & $\bullet$ & $\circ$ & $\circ$ & $\bullet$ & $\circ$ & $\bullet$ & $\circ$ & $\bullet$ & $\bullet$ & 2048	& $\bullet$ & $\bullet$ & $\bullet$ & $\circ$ & $\bullet$ \\
Apache & 	14.04 & 	1.0.1f & 	C	& 1 	& $\circ$ & $\bullet$ & $\bullet$ & $\bullet$ & $\bullet$ & $\circ$ & $\circ$ & $\bullet$ & $\bullet$ & $\circ$ & $\circ$ & $\circ$ & $\circ$ & $\circ$ & $\bullet$ & $\circ$ & $\circ$ & $\circ$ & $\bullet$ & $\bullet$ & 2048	& $\bullet$ & $\bullet$ & $\bullet$ & $\circ$ & $\bullet$ \\
Apache & 	16.04 & 	1.0.2g	 & B & 	78   	& $\circ$ & $\bullet$ & $\bullet$ & $\bullet$ & $\circ$ & $\circ$ & $\circ$ & $\bullet$ & $\bullet$ & $\circ$ & $\circ$ & $\circ$ & $\circ$ & $\circ$ & $\bullet$ & $\circ$ & $\circ$ & $\circ$ & $\bullet$ & $\bullet$ & 2048	& $\bullet$ & $\bullet$ & $\bullet$ & $\circ$ & $\bullet$ \\
Apache	 & 18.04 & 	1.1.0g$^2$	 & B & 	15  	& $\circ$ & $\bullet$ & $\bullet$ & $\bullet$ & $\circ$ & $\circ$ & $\circ$ & $\bullet$ & $\bullet$ & $\circ$ & $\circ$ & $\circ$ & $\circ$ & $\bullet$ & $\bullet$ & $\circ$ & $\circ$ & $\bullet$ & $\bullet$ & $\bullet$ & 2048	& $\bullet$ & $\bullet$ & $\bullet$ & $\circ$ 	& $\bullet$ \\
Apache & 	18.04 & 	1.1.0g$^2$	 & B & 	4   	& $\bullet$ & $\bullet$ & $\bullet$ & $\bullet$ & $\circ$ & $\circ$ & $\circ$ & $\bullet$ & $\bullet$ & $\circ$ & $\circ$ & $\circ$ & $\circ$ & $\bullet$ & $\bullet$ & $\circ$ & $\circ$ & $\bullet$ & $\bullet$ & $\bullet$ & 2048	& $\bullet$ & $\bullet$ & $\bullet$ & $\circ$ 	& $\bullet$ \\
\bottomrule
\end{tabular*}

%% file: 7_recommendations.tex
\subsection{Online Recommendations}

With only a fraction of sites matching defaults and with defaults receiving Bs
or lower, we turn our attention to HTTPS configuration guides to understand how
operators try to securely configure sites. We collect a sample of likely-found
recommendations by having four researchers independently compile a list of
relevant search terms for Nginx and Apache, and then analyze the configuration
guides from the first three pages of Google results for each search. In total,
the team generated 16~unique search terms for Nginx and 9~for Apache e.g.,
``secure SSL configuration apache'', ``secure nginx ssl settings''.). In cases
where websites recommended multiple configurations (e.g., for modern and legacy,
or wide compatibility), we extracted all recommendations.  To ensure our results
are not biased towards English recommendations, we also searched for Nginx TLS
recommendations in the next top three languages (after English): German,
Russian, and French.  These languages correspond to the countries where the
highest number of non-Cloudflare sites geolocated.  We performed our searches
over ExpressVPN using the country-specific Google search engine for each.
Another concern is that the exact recommendations on each site may have changed
over time and what we observe today may not be reflective of what server
operators may have seen when they were setting up web servers, potentially
several years ago.  For this reason, we also compiled a list of historical
recommendations.  Specifically, we selected a random sample of 50 unique URLs
for which we inspected historical snapshots from WayBack
Machine~\cite{wayback}---a free online service that stores snapshots of websites
over time.

In total, we collected 1,162~search results, from which we extracted 466~unique
recommendation sites, including 57~URLs in French, 76~URLs in German, and
58~URLs in Russian. About 40\% of sites do not contain a recommendation. These
sites tend to be websites related to SSL configurations but do not contain a
specific, ready-to-use recommendation. For example, a large number of sites
cover topics related to installing SSL certificate or setting up an SSL proxy.
Usable recommendations typically specify a handful of settings (e.g., version,
ciphers, and server cipher preference), but most do not account for all
configurable options.  About 26\% of recommendations are from commercial sites
(e.g., official documentation and cloud tutorials), 63\% are blogs (73\% of
which are personal and 27\% corporate), and 12\% are forum posts on sites like
Stack Overflow.  The highest ranked URLs tend to be official sources.  The SSL
Mozilla configuration generator and an official Nginx documentation page show up
most frequently in search results, followed by posts from TechRepublic,
DigitalOcean, CyberCiti, Apache, Linode, TechMint, and GeekFlare. The first
non-official URL is ``haydenjames.io'', which appears in results from 6 of the
38 search terms.

Measuring which websites follow a specific recommendation is complicated by the
fact that applying recommendations to different versions of Apache, Nginx, and
OpenSSL can lead to different TLS feature support. First, different versions of
OpenSSL support different TLS versions, cipher suites, and extensions. Second,
recommendations of SSL protocol versions and ciphers are often stated in terms
of exclusions (e.g., \texttt{All -SSLv2}) and components (e.g.,
\texttt{ECDHE+AESGCM}). For this reason, we check \textit{consistency} with
recommendations and check for support of allowed or disallowed options and
cipher components instead of the exact cipher strings. For example, given the
recommendation \texttt{ECDHE+AESGCM:!RC4}, a configuration would be consistent
with it if at least one \texttt{ECDHE+AESGCM} cipher suite is supported, no
non-\texttt{ECDHE+AESGCM} cipher suite is supported, including no RC4 cipher
suites. We thus offer an upper bound estimate on the number of sites that follow
a recommendation. We limit our analysis to non-Cloudflare sites hosted on Apache
and Nginx. We find that 4\% of Apache sites and 20\% of Nginx sites in Tranco
are consistent with a recommendation.  Of sites that identify themselves as
running on Ubuntu, 18\% of Apache sites are consistent with a recommendation and
7.3\% of Nginx sites are consistent.
Because we know the default values of Ubuntu, we can calculate the exact options
for each recommendation by applying them to the defaults from all versions of
Apache, Nginx, and OpenSSL on Ubuntu 14.04--18.04. In the case of a
recommendation that is missing directives, we substitute the default setting.
We find almost no matches when searching for exact recommendations in our
data---under 1\% of both Tranco and CT sites. While many sites running Ubuntu
likely do not include the OS in the \texttt{Server} header, the result confirms
that our \emph{consistent} metric significantly overestimates the number of
sites that follow a recommendation and should only be treated as a high upper
bound.

\paragraph{Security of online recommendations} We apply our grading scheme to
the recommendations we collect to understand how they would affect server
security.  Interestingly, only 7.3\% of unique recommendations receive a grade
of A.  The large majority---88\%, receive either a grade of B or C, and 4.0\%
receive an F.  We find that secure recommendations rank higher than insecure
recommendations.  The average rank for recommendations receiving an A is 9.8,
versus 14.4 for B and 12.6 for C (differences are statistically significant
using an ANOVA test with a significance level 0.05, p$<$0.001).

In terms of individual TLS parameter support, several findings stand out.
First, only a small proportion of recommendations would result in a modern,
A-grade configuration with only the most up-to-date security parameters (only
22\% recommend TLSv1.3, 30\% recommend ChaCha ciphers).  If applied to the
default configuration of a server, many would not exclude insecure options.
Specifically, 8\% of recommendations support SSLv3, 3\% support DES, 18\%
support 3DES, 13\% support RC4, and 8\% support MD5.  More recommendations still
allow ciphers that are not widely used (38\% Camellia, 16\% SEED and ARIA
ciphers).

There are several options that are consistently recommended across the majority
of sources.  Just under a third of recommendations  mention session tickets
(31\%).  Where mentioned, few recommendations suggest that session tickets be
turned on (11\% of current).  This is contrary to the wide support that we
observe empirically---74\% of non-Cloudflare domains support session tickets in
Tranco.  Another option where recommendations are generally unanimous is
\texttt{ssl\_prefer\_server\_ciphers}.  This directive is mentioned in 75\% of
recommendations, 91\% of which recommend that it be turned on.  This is
higher than the level of support we empirically observe in Tranco (63\%).
Historical recommendations follow a qualitatively similar patterns and are
omitted for brevity.  We do not find significant differences between current and
historical recommendations in terms of security posture, and they do not result
in a higher proportion of matches.

One possible explanation for observing such a larger number of insecure
recommendations may be that they are outdated.  The majority of sites display
the publication date (77\%), and we investigate whether the publication date has
any impact on the security posture of the recommendation.  Despite a long tail
of publication years dating back to 2010, we find that 39\% of recommendations
are ``current'', meaning that they were published on or after January 1, 2019.
Of sites that are current, 46\% are secure, versus 56\% of secure sites
published prior to 2019.  A Chi-square test confirms that there is no
statistically significant relationship between the security posture of a
recommendation (secure versus not) and publication date (current versus not),
(d.f.=1, p = 3.0112, p = 0.0827).  This result points to a worrying trend that
the security posture of recommendations is not improving over time.  This may be
in part due to the fundamental nature of the TLS protocol, which makes it such
that servers wishing to support older clients need to adopt legacy
recommendations that are inevitably less secure.

These results highlight a surprising amount of variance and a high degree of
inconsistency across recommendations---a pattern that has been observed in other
areas of security as well~\cite{reeder2017152,redmiles255292}.  Moreover, they
do not consistently point users to the most secure configurations.  This may be
in part, because server operators must also consider backwards compatibility.
Nevertheless, approximately 10\% of recommendations still allow MD5 MACs and RC4
ciphers, neither of which are needed to support the modern browsers (Chrome,
Safari, Edge, Firefox).

%% file: 8_discussion.tex
\section{Discussion}

We initially expected to find large clusters of sites using
the same configurations. We hoped that we would be able to match these common
configurations to online recommendation guides, Stack Overflow posts, blogs,
server defaults, and administration tools like Certbot~\cite{certbot}. Instead,
we find a complex and fragmented ecosystem of HTTPS configurations with a long tail of individually configured sites. Our attempts to cluster configurations and find commonalities were
largely unsuccessful---there are simply a large number of unique, seemingly
unrelated configurations deployed.
In nearly all cases, when a large number of sites use the
same configuration, it is due to SaaS providers deploying the same configuration
on all sites.  Despite the complexity of the ecosystem,
there are several important high-level takeaways for the security community.

Our results highlight the importance of doing analysis at the configuration-level, 
as opposed to adopting a uni-dimensional view that considers support for an
individual directive at a time. Adopting such a uni-dimensional view may
mask patterns that only emerge at the configuration-level, such as the influence
of geography, cloud providers, provider defaults, and online recommendations.
We find that using a cloud provider is one of the strongest predictors for a site being configured correctly. Most individually managed servers struggle to achieve secure configurations.
It is an important distinction that operators are not getting better at configuring servers, but are instead migrating to platforms where they can sidestep the process.  
Although we cannot quantify exactly how much SaaS providers pull up 
the overall security of the internet, we provide a lower bound by manually labeling configurations from the largest and most well-known SaaS providers, and show that 
it is significant enough in order for security researchers to distinguish it 
when evaluating the security posture of the overall internet.
This may be relevant for a number of research directions: 
when the aim is to understand the decision-making process of server operators,
when evaluating the adoption of new security features or technologies, 
as well as the overall security posture of the internet.

It is unclear why we find such little evidence of online recommendations being used in practice.
It may be because very few server operators use online recommendations exactly as specified,
or another component of the ecosystem that we do not observe is contributing to the process.
For example, server operators may individually remove cipher suites over time as
vulnerabilities are announced, they may use online grading tools such as SSL Labs and
edit their recommendations to obtain good grades, or they may use additional software
that automatically configures TLS (e.g., Certbot). Our results may also add important 
context to endeavors to invest in tools and resources to improve online recommendations, 
such as a recent effort from the Internet Society (ISOC), who is actively investing in 
and developing recommendations for a variety of server software~\cite{isoc_announcement}. 

Lastly, our results point to the importance of secure defaults that do not require
operators to seek out trustworthy sources for how to securely configure
sites post-installation---a system which appears to be fundamentally broken. Echoing previous
studies~\cite{krombholz2017,bernhard2019use}, we encourage the security community 
to build future systems  that prioritize being secure ``out of the box''.

%% file: 10_conclusion.tex
\section{Conclusion}

We find a complex and fragmented ecosystem of HTTPS configurations with a long
tail of individually configured websites. While the majority of sites 
are securely configured, this is primarily due to cloud providers,
which provide secure configurations for all hosted
websites. Using a cloud provider is one of the strongest predictors for a site
being configured correctly. Most individually managed servers, however, continue
to struggle to achieve secure configurations. Few operators use server defaults
for Nginx and Apache, and there is a lack of adherence to online recommendations,
which may be in part because these are often incomplete and insecure. 
Our findings point to the need for future analysis of  HTTPS security to take into account 
these two disparate ecosystems, and for the security community to invest in solutions 
that bridge the security gap between the two.

%% file: appendix.tex
\appendix

%
%
%

\onecolumn
\begin{table}[t]
\footnotesize
\centering

\input{tables/saas_defaults.tex}

\caption{
The five most dominant configuration from the top 10 most popular ASs in terms
of the proportion of sites hosted on the AS (\% AS).}
\label{tab:saas_defaults}
\end{table}

\onecolumn
\begin{table}[t]
\footnotesize
\centering
\input{tables/appendix_stats.tex}
\caption{Support for select TLS configuration components by server. 
Each value represents the proportion of domains in each server within Tranco and CT, respectively.}
\label{tab:binary_feature_stats}
\end{table}

\onecolumn
\begin{table}[t]
\footnotesize
\centering
\input{tables/appendix_model.tex}
\caption{Generalized ordinal model results in terms of marginal odds for each grade category (F/C, B, and A).
Signif. codes:  0 *** 0.001, ** 0.01, * 0.05, . 0.1}
\label{tab:complete_model_results}
\end{table}

\section{Search terms for recommendations}

The search terms used to search for recommendations (n=26 English terms) are as follows:
secure ssl configuration nginx*,  
secure ssl configuration apache,
nginx set up tls,
apache set up tls,
secure apache ssl settings,
secure nginx ssl settings*,
best nginx ssl config,
best apache ssl config,
ssllabs a plus apache,
ssllabs a plus nginx,
configure SSL nginx*,
nginx https,
nginx https configuration,
nginx conf file ssl settings*,
nginx secure configuration,
nginx secure ssl configuration,
nginx A ssl labs,
nginx secure ciphers,
disable weak ciphers nginx,
nginx ssl configuration,
nginx configure tls,
apache config tls,
apache set up https,
setup apache secure tls configuration,
harden tls configuration apache web server.
Four of the search terms were translated into French, German, and Russian, for a total of 38 search terms (asterisked terms were translated).

%% file: tables/saas_defaults.tex

\begin{tabular*}{\textwidth}{l%
	l@{\hskip -8pt}%
	l@{\hskip -2pt}%
	l@{\hskip -8pt}%
	l@{\hskip -8pt}%
	l@{\hskip -8pt}%
	l@{\hskip -8pt}%
	l@{\hskip -8pt}%
	l@{\hskip -8pt}%
	l@{\hskip -8pt}%
	l@{\hskip -8pt}%
	l@{\hskip -8pt}%
	l@{\hskip -8pt}%
	l@{\hskip -8pt}%
	l@{\hskip -8pt}%
	l@{\hskip -8pt}%
	l@{\hskip -8pt}%
	l@{\hskip -8pt}%
	l@{\hskip -8pt}%
	l@{\hskip -8pt}%
	l@{\hskip -8pt}%
	l@{\hskip -8pt}%
	l@{\hskip -8pt}%
	l@{\hskip -8pt}%
	l@{\hskip -8pt}%
	l@{\hskip -8pt}%
	l@{\hskip -8pt}%
	l@{\hskip -8pt}%
	l@{\hskip -8pt}%
	l@{\hskip -8pt}%
	l@{\hskip -8pt}%
	llllccrrrrrrrrrrrrrrrrrrrrrcrrrrr}
\toprule
\begin{turn}{65} AS Name $\;\;\;\;\;\;\;\;$ \end{turn} &
\begin{turn}{65} Num Sites $\;\;\;\;\;\;\;\;\;\;\;\;\;\;\;\;\;\;\;\;\;$ \end{turn} &
\begin{turn}{65} \% AS $\;\;\;\;\;\;\;\;\;\;\;\;\;$ \end{turn} &
\begin{turn}{65} Grade $\;\;\;\;\;\;\;\;\;\;\;\;\;$ \end{turn} &
\begin{turn}{65} TSLv1.3 $\;\;\;\;\;\;\;\;\;\;\;$ \end{turn} &
\begin{turn}{65} TSLv1.2 $\;\;\;\;\;\;\;\;\;\;\;$ \end{turn}  &
\begin{turn}{65} TSLv1.1 $\;\;\;\;\;\;\;\;\;\;\;$ \end{turn} &
\begin{turn}{65} TSLv1.0 $\;\;\;\;\;\;\;\;\;\;\;$ \end{turn} &
\begin{turn}{65} SSLv3.0 $\;\;\;\;\;\;\;\;\;\;\;$ \end{turn} &
\begin{turn}{65} Server pref $\;\;\;\;\;\;\;\;$ \end{turn} &
\begin{turn}{65} Sesssion ID $\;\;\;\;\;\;\;\;$ \end{turn} &
\begin{turn}{65} Session Ticket   \end{turn} &
\begin{turn}{65} RC4 $\;\;\;\;\;\;\;\;\;\;\;\;\;$ \end{turn} &
\begin{turn}{65} DES $\;\;\;\;\;\;\;\;\;\;$ \end{turn}&
\begin{turn}{65} 3DES $\;\;\;\;\;\;\;\;\;\;\;\;\;\;$ \end{turn} &
\begin{turn}{65} Export $\;\;\;\;\;\;\;\;\;\;$ \end{turn}&
\begin{turn}{65} ARIA $\;\;\;\;\;\;\;\;\;\;$ \end{turn}&
\begin{turn}{65} Camellia $\;\;\;\;\;\;\;\;\;$ \end{turn}&
\begin{turn}{65} IDEA $\;\;\;\;\;\;\;\;\;\;$ \end{turn}&
\begin{turn}{65} SEED $\;\;\;\;\;\;\;\;\;\;\;\;\;\;$ \end{turn}&
\begin{turn}{65} CHACHA  $\;\;\;\;\;\;\;\;\;$ \end{turn}&
\begin{turn}{65} AEAD $\;\;\;\;\;\;\;\;\;\;\;\;\;$ \end{turn}&
\begin{turn}{65} AES-GCM $\;\;\;\;\;\;\;\;\;\;\;\;\;$ \end{turn}&
\begin{turn}{65} DH Group $\;\;\;\;\;\;\;\;\;\;\;\;\;\;$ \end{turn}&
\begin{turn}{65} KE:DHE $\;\;\;\;\;\;\;\;\;\;\;\;\;$ \end{turn}&
\begin{turn}{65} KE:ECDHE $\;\;\;\;\;\;\;\;\;\;\;$ \end{turn}&
\begin{turn}{65} KE:RSA $\;\;\;\;\;\;\;\;\;\;\;$ \end{turn}&
\begin{turn}{65} MD5 $\;\;\;\;\;\;\;\;\;\;\;\;$  \end{turn}&
\begin{turn}{65} SHA  \end{turn} \\
\midrule
Cloudflare & 	93,083	&	70	&	A	&	$\bullet$	&	$\bullet$	&	$\bullet$	&	$\bullet$	&	$\circ$	&	$\circ$	&	$\bullet$	&	$\bullet$	&	$\circ$	&	$\circ$	&	$\circ$	&	$\circ$	&	$\circ$	&	$\circ$	&	$\circ$	&	$\circ$	&	$\bullet$	&	$\bullet$	&	$\bullet$	&	-	&	$\circ$	&	$\bullet$	&	$\circ$	&	$\circ$	&	$\bullet$	\\
Cloudflare &	22,123	&	16.6	&	A	&	$\bullet$	&	$\bullet$	&	$\bullet$	&	$\bullet$	&	$\circ$	&	$\circ$	&	$\bullet$	&	$\bullet$	&	$\circ$	&	$\circ$	&	$\circ$	&	$\circ$	&	$\circ$	&	$\circ$	&	$\circ$	&	$\circ$	&	$\bullet$	&	$\bullet$	&	$\bullet$	&	-	&	$\circ$	&	$\bullet$	&	$\bullet$	&	$\circ$	&	$\bullet$	\\
Cloudflare 	& 8,771	&	6.6	&	A	&	$\bullet$	&	$\bullet$	&	$\circ$	&	$\circ$	&	$\circ$	&	$\circ$	&	$\bullet$	&	$\bullet$	&	$\circ$	&	$\circ$	&	$\circ$	&	$\circ$	&	$\circ$	&	$\circ$	&	$\circ$	&	$\circ$	&	$\bullet$	&	$\bullet$	&	$\bullet$	&	-	&	$\circ$	&	$\bullet$	&	$\bullet$	&	$\circ$	&	$\bullet$	\\
Cloudflare 	& 2,276	&	1.7	&	A	&	$\circ$	&	$\bullet$	&	$\bullet$	&	$\bullet$	&	$\circ$	&	$\circ$	&	$\bullet$	&	$\bullet$	&	$\circ$	&	$\circ$	&	$\circ$	&	$\circ$	&	$\circ$	&	$\circ$	&	$\circ$	&	$\circ$	&	$\bullet$	&	$\bullet$	&	$\bullet$	&	-	&	$\circ$	&	$\bullet$	&	$\bullet$	&	$\circ$	&	$\bullet$	\\
Cloudflare &	1,588	&	1.2	&	A	&	$\bullet$	&	$\bullet$	&	$\circ$	&	$\circ$	&	$\circ$	&	$\circ$	&	$\bullet$	&	$\bullet$	&	$\circ$	&	$\circ$	&	$\circ$	&	$\circ$	&	$\circ$	&	$\circ$	&	$\circ$	&	$\circ$	&	$\bullet$	&	$\bullet$	&	$\bullet$	&	-	&	$\circ$	&	$\bullet$	&	$\circ$	&	$\circ$	&	$\bullet$	\\
\midrule
Amazon-02  	&	5,441	&	14.2	&	A	&	$\circ$	&	$\bullet$	&	$\bullet$	&	$\circ$	&	$\circ$	&	$\bullet$	&	$\circ$	&	$\bullet$	&	$\circ$	&	$\circ$	&	$\circ$	&	$\circ$	&	$\circ$	&	$\circ$	&	$\circ$	&	$\circ$	&	$\circ$	&	$\bullet$	&	$\bullet$	&	-	&	$\circ$	&	$\bullet$	&	$\bullet$	&	$\circ$	&	$\bullet$	\\
Amazon-02  	&	4,563	&	11.9	&	A	&	$\circ$	&	$\bullet$	&	$\bullet$	&	$\bullet$	&	$\circ$	&	$\bullet$	&	$\bullet$	&	$\bullet$	&	$\circ$	&	$\circ$	&	$\circ$	&	$\circ$	&	$\circ$	&	$\circ$	&	$\circ$	&	$\circ$	&	$\circ$	&	$\bullet$	&	$\bullet$	&	-	&	$\circ$	&	$\bullet$	&	$\bullet$	&	$\circ$	&	$\bullet$	\\
Amazon-02  	&	3,074	&	8.0	&	A	&	$\circ$	&	$\bullet$	&	$\bullet$	&	$\bullet$	&	$\circ$	&	$\bullet$	&	$\circ$	&	$\bullet$	&	$\circ$	&	$\circ$	&	$\bullet$	&	$\circ$	&	$\circ$	&	$\circ$	&	$\circ$	&	$\circ$	&	$\circ$	&	$\bullet$	&	$\bullet$	&	-	&	$\circ$	&	$\bullet$	&	$\bullet$	&	$\circ$	&	$\bullet$	\\
Amazon-02  	&	1,696	&	4.4	&	A	&	$\circ$	&	$\bullet$	&	$\bullet$	&	$\bullet$	&	$\circ$	&	$\bullet$	&	$\bullet$	&	$\circ$	&	$\circ$	&	$\circ$	&	$\circ$	&	$\circ$	&	$\circ$	&	$\circ$	&	$\circ$	&	$\circ$	&	$\circ$	&	$\bullet$	&	$\bullet$	&	-	&	$\circ$	&	$\bullet$	&	$\bullet$	&	$\circ$	&	$\bullet$	\\
Amazon-02  	&	1,100	&	2.9	&	B	&	$\circ$	&	$\bullet$	&	$\bullet$	&	$\bullet$	&	$\circ$	&	$\bullet$	&	$\bullet$	&	$\bullet$	&	$\circ$	&	$\circ$	&	$\bullet$	&	$\circ$	&	$\circ$	&	$\circ$	&	$\circ$	&	$\circ$	&	$\circ$	&	$\bullet$	&	$\bullet$	&	2048	&	$\bullet$	&	$\bullet$	&	$\bullet$	&	$\circ$	&	$\bullet$	\\
\midrule
Google	&	11,978	&	38.6	&	B	&	$\circ$	&	$\bullet$	&	$\bullet$	&	$\circ$	&	$\circ$	&	$\bullet$	&	$\bullet$	&	$\bullet$	&	$\circ$	&	$\circ$	&	$\circ$	&	$\circ$	&	$\circ$	&	$\circ$	&	$\circ$	&	$\circ$	&	$\circ$	&	$\bullet$	&	$\bullet$	&	2048	&	$\bullet$	&	$\bullet$	&	$\bullet$	&	$\circ$	&	$\bullet$	\\
Google	&	2,461	&	7.9	&	A	&	$\circ$	&	$\bullet$	&	$\bullet$	&	$\bullet$	&	$\circ$	&	$\bullet$	&	$\circ$	&	$\bullet$	&	$\circ$	&	$\circ$	&	$\bullet$	&	$\circ$	&	$\circ$	&	$\circ$	&	$\circ$	&	$\circ$	&	$\bullet$	&	$\bullet$	&	$\bullet$	&	-	&	$\circ$	&	$\bullet$	&	$\bullet$	&	$\circ$	&	$\bullet$	\\
Google	&	2,028	&	6.5	&	B	&	$\bullet$	&	$\bullet$	&	$\bullet$	&	$\bullet$	&	$\circ$	&	$\circ$	&	$\circ$	&	$\circ$	&	$\circ$	&	$\circ$	&	$\bullet$	&	$\circ$	&	$\circ$	&	$\circ$	&	$\circ$	&	$\circ$	&	$\bullet$	&	$\bullet$	&	$\bullet$	&	-	&	$\circ$	&	$\bullet$	&	$\bullet$	&	$\circ$	&	$\bullet$	\\
Google	&	1,790	&	5.8	&	B	&	$\circ$	&	$\bullet$	&	$\bullet$	&	$\bullet$	&	$\circ$	&	$\circ$	&	$\circ$	&	$\bullet$	&	$\circ$	&	$\circ$	&	$\bullet$	&	$\circ$	&	$\circ$	&	$\circ$	&	$\circ$	&	$\circ$	&	$\bullet$	&	$\bullet$	&	$\bullet$	&	-	&	$\circ$	&	$\bullet$	&	$\bullet$	&	$\circ$	&	$\bullet$	\\
Google	&	1,742	&	5.6	&	A	&	$\bullet$	&	$\bullet$	&	$\bullet$	&	$\circ$	&	$\circ$	&	$\bullet$	&	$\circ$	&	$\bullet$	&	$\circ$	&	$\circ$	&	$\circ$	&	$\circ$	&	$\circ$	&	$\circ$	&	$\circ$	&	$\circ$	&	$\bullet$	&	$\bullet$	&	$\bullet$	&	-	&	$\circ$	&	$\bullet$	&	$\bullet$	&	$\circ$	&	$\bullet$	\\
\midrule
Unified Layer	&	15,519	&	57.1	&	B	&	$\circ$	&	$\bullet$	&	$\bullet$	&	$\bullet$	&	$\circ$	&	$\bullet$	&	$\bullet$	&	$\bullet$	&	$\circ$	&	$\circ$	&	$\circ$	&	$\circ$	&	$\circ$	&	$\circ$	&	$\circ$	&	$\circ$	&	$\circ$	&	$\bullet$	&	$\bullet$	&	2048	&	$\bullet$	&	$\bullet$	&	$\circ$	&	$\circ$	&	$\bullet$	\\
Unified Layer	&	5,008	&	18.4	&	B	&	$\circ$	&	$\bullet$	&	$\bullet$	&	$\bullet$	&	$\circ$	&	$\bullet$	&	$\bullet$	&	$\bullet$	&	$\circ$	&	$\circ$	&	$\circ$	&	$\circ$	&	$\bullet$	&	$\circ$	&	$\circ$	&	$\circ$	&	$\bullet$	&	$\bullet$	&	$\bullet$	&	-	&	$\circ$	&	$\bullet$	&	$\circ$	&	$\circ$	&	$\bullet$	\\
Unified Layer	&	958	&	3.5	&	B	&	$\circ$	&	$\bullet$	&	$\circ$	&	$\circ$	&	$\circ$	&	$\circ$	&	$\bullet$	&	$\bullet$	&	$\circ$	&	$\circ$	&	$\circ$	&	$\circ$	&	$\circ$	&	$\circ$	&	$\circ$	&	$\circ$	&	$\circ$	&	$\bullet$	&	$\bullet$	&	-	&	$\circ$	&	$\bullet$	&	$\circ$	&	$\circ$	&	$\bullet$	\\
Unified Layer	&	764	&	2.8	&	C	&	$\circ$	&	$\bullet$	&	$\bullet$	&	$\bullet$	&	$\circ$	&	$\circ$	&	$\bullet$	&	$\bullet$	&	$\bullet$	&	$\circ$	&	$\bullet$	&	$\circ$	&	$\circ$	&	$\bullet$	&	$\bullet$	&	$\bullet$	&	$\circ$	&	$\bullet$	&	$\bullet$	&	-	&	$\circ$	&	$\bullet$	&	$\bullet$	&	$\bullet$	&	$\bullet$	\\
Unified Layer	&	500	&	1.8	&	B	&	$\circ$	&	$\bullet$	&	$\bullet$	&	$\bullet$	&	$\circ$	&	$\bullet$	&	$\bullet$	&	$\bullet$	&	$\circ$	&	$\circ$	&	$\bullet$	&	$\circ$	&	$\circ$	&	$\bullet$	&	$\circ$	&	$\circ$	&	$\circ$	&	$\bullet$	&	$\bullet$	&	2048	&	$\bullet$	&	$\bullet$	&	$\bullet$	&	$\circ$	&	$\bullet$	\\
\midrule
Amazon AES	&	3,270	&	14.2	&	A	&	$\circ$	&	$\bullet$	&	$\bullet$	&	$\bullet$	&	$\circ$	&	$\bullet$	&	$\bullet$	&	$\bullet$	&	$\circ$	&	$\circ$	&	$\circ$	&	$\circ$	&	$\circ$	&	$\circ$	&	$\circ$	&	$\circ$	&	$\circ$	&	$\bullet$	&	$\bullet$	&	-	&	$\circ$	&	$\bullet$	&	$\bullet$	&	$\circ$	&	$\bullet$	\\
Amazon AES	&	1,522	&	6.6	&	A	&	$\circ$	&	$\bullet$	&	$\bullet$	&	$\bullet$	&	$\circ$	&	$\bullet$	&	$\bullet$	&	$\circ$	&	$\circ$	&	$\circ$	&	$\circ$	&	$\circ$	&	$\circ$	&	$\circ$	&	$\circ$	&	$\circ$	&	$\circ$	&	$\bullet$	&	$\bullet$	&	-	&	$\circ$	&	$\bullet$	&	$\bullet$	&	$\circ$	&	$\bullet$	\\
Amazon AES	&	1,494	&	6.5	&	B	&	$\circ$	&	$\bullet$	&	$\bullet$	&	$\bullet$	&	$\circ$	&	$\bullet$	&	$\bullet$	&	$\bullet$	&	$\circ$	&	$\circ$	&	$\bullet$	&	$\circ$	&	$\circ$	&	$\circ$	&	$\circ$	&	$\circ$	&	$\circ$	&	$\bullet$	&	$\bullet$	&	2048	&	$\bullet$	&	$\bullet$	&	$\bullet$	&	$\circ$	&	$\bullet$	\\
Amazon AES	&	1,022	&	4.4	&	A	&	$\circ$	&	$\bullet$	&	$\bullet$	&	$\circ$	&	$\circ$	&	$\bullet$	&	$\bullet$	&	$\bullet$	&	$\circ$	&	$\circ$	&	$\circ$	&	$\circ$	&	$\circ$	&	$\circ$	&	$\circ$	&	$\circ$	&	$\circ$	&	$\bullet$	&	$\bullet$	&	-	&	$\circ$	&	$\bullet$	&	$\bullet$	&	$\circ$	&	$\bullet$	\\
Amazon AES	&	791	&	3.4	&	A	&	$\circ$	&	$\bullet$	&	$\circ$	&	$\circ$	&	$\circ$	&	$\bullet$	&	$\bullet$	&	$\bullet$	&	$\circ$	&	$\circ$	&	$\circ$	&	$\circ$	&	$\circ$	&	$\circ$	&	$\circ$	&	$\circ$	&	$\circ$	&	$\bullet$	&	$\bullet$	&	-	&	$\circ$	&	$\bullet$	&	$\bullet$	&	$\circ$	&	$\bullet$	\\
\midrule
OVH	&	2,943	&	13.9	&	B	&	$\circ$	&	$\bullet$	&	$\bullet$	&	$\bullet$	&	$\circ$	&	$\bullet$	&	$\circ$	&	$\circ$	&	$\circ$	&	$\circ$	&	$\circ$	&	$\circ$	&	$\circ$	&	$\circ$	&	$\circ$	&	$\circ$	&	$\circ$	&	$\bullet$	&	$\bullet$	&	2048	&	$\bullet$	&	$\bullet$	&	$\bullet$	&	$\circ$	&	$\bullet$	\\
OVH	&	2,034	&	9.6	&	B	&	$\circ$	&	$\bullet$	&	$\circ$	&	$\circ$	&	$\circ$	&	$\circ$	&	$\bullet$	&	$\bullet$	&	$\circ$	&	$\circ$	&	$\circ$	&	$\circ$	&	$\circ$	&	$\circ$	&	$\circ$	&	$\circ$	&	$\circ$	&	$\bullet$	&	$\bullet$	&	-	&	$\circ$	&	$\bullet$	&	$\circ$	&	$\circ$	&	$\bullet$	\\
OVH	&	749	&	3.5	&	B	&	$\circ$	&	$\bullet$	&	$\bullet$	&	$\bullet$	&	$\circ$	&	$\bullet$	&	$\circ$	&	$\bullet$	&	$\circ$	&	$\circ$	&	$\circ$	&	$\circ$	&	$\circ$	&	$\bullet$	&	$\circ$	&	$\circ$	&	$\circ$	&	$\bullet$	&	$\bullet$	&	-	&	$\circ$	&	$\bullet$	&	$\bullet$	&	$\circ$	&	$\bullet$	\\
OVH	&	705	&	3.3	&	B	&	$\circ$	&	$\bullet$	&	$\bullet$	&	$\bullet$	&	$\circ$	&	$\circ$	&	$\bullet$	&	$\bullet$	&	$\circ$	&	$\circ$	&	$\bullet$	&	$\circ$	&	$\circ$	&	$\circ$	&	$\circ$	&	$\circ$	&	$\circ$	&	$\bullet$	&	$\bullet$	&	2048	&	$\bullet$	&	$\bullet$	&	$\bullet$	&	$\circ$	&	$\bullet$	\\
OVH	&	690	&	3.3	&	B	&	$\circ$	&	$\bullet$	&	$\bullet$	&	$\bullet$	&	$\circ$	&	$\bullet$	&	$\bullet$	&	$\circ$	&	$\circ$	&	$\circ$	&	$\circ$	&	$\circ$	&	$\circ$	&	$\circ$	&	$\circ$	&	$\circ$	&	$\circ$	&	$\bullet$	&	$\bullet$	&	2048	&	$\bullet$	&	$\bullet$	&	$\bullet$	&	$\circ$	&	$\bullet$	\\
\midrule
Hetzner	&	1,712	&	9.4	&	B	&	$\circ$	&	$\bullet$	&	$\circ$	&	$\circ$	&	$\circ$	&	$\circ$	&	$\bullet$	&	$\bullet$	&	$\circ$	&	$\circ$	&	$\circ$	&	$\circ$	&	$\circ$	&	$\circ$	&	$\circ$	&	$\circ$	&	$\circ$	&	$\bullet$	&	$\bullet$	&	-	&	$\circ$	&	$\bullet$	&	$\circ$	&	$\circ$	&	$\bullet$	\\
Hetzner	&	1,017	&	5.6	&	B	&	$\circ$	&	$\bullet$	&	$\bullet$	&	$\bullet$	&	$\circ$	&	$\bullet$	&	$\bullet$	&	$\bullet$	&	$\circ$	&	$\circ$	&	$\circ$	&	$\circ$	&	$\circ$	&	$\bullet$	&	$\circ$	&	$\circ$	&	$\circ$	&	$\bullet$	&	$\bullet$	&	2048	&	$\bullet$	&	$\bullet$	&	$\bullet$	&	$\circ$	&	$\bullet$	\\
Hetzner	&	640	&	3.5	&	A	&	$\circ$	&	$\bullet$	&	$\circ$	&	$\circ$	&	$\circ$	&	$\bullet$	&	$\bullet$	&	$\bullet$	&	$\circ$	&	$\circ$	&	$\circ$	&	$\circ$	&	$\circ$	&	$\circ$	&	$\circ$	&	$\circ$	&	$\bullet$	&	$\bullet$	&	$\bullet$	&	-	&	$\circ$	&	$\bullet$	&	$\circ$	&	$\circ$	&	$\bullet$	\\
Hetzner	&	546	&	3.0	&	B	&	$\circ$	&	$\bullet$	&	$\bullet$	&	$\bullet$	&	$\circ$	&	$\bullet$	&	$\bullet$	&	$\bullet$	&	$\circ$	&	$\circ$	&	$\bullet$	&	$\circ$	&	$\circ$	&	$\circ$	&	$\circ$	&	$\circ$	&	$\circ$	&	$\bullet$	&	$\bullet$	&	2048	&	$\bullet$	&	$\bullet$	&	$\bullet$	&	$\circ$	&	$\bullet$	\\
Hetzner	&	539	&	3.0	&	B	&	$\circ$	&	$\bullet$	&	$\bullet$	&	$\bullet$	&	$\circ$	&	$\bullet$	&	$\circ$	&	$\bullet$	&	$\circ$	&	$\circ$	&	$\circ$	&	$\circ$	&	$\circ$	&	$\bullet$	&	$\circ$	&	$\circ$	&	$\circ$	&	$\bullet$	&	$\bullet$	&	-	&	$\circ$	&	$\bullet$	&	$\bullet$	&	$\circ$	&	$\bullet$	\\
\midrule
GoDaddy	&	8,101	&	51.7	&	B	&	$\circ$	&	$\bullet$	&	$\bullet$	&	$\circ$	&	$\circ$	&	$\bullet$	&	$\bullet$	&	$\bullet$	&	$\circ$	&	$\circ$	&	$\bullet$	&	$\circ$	&	$\circ$	&	$\bullet$	&	$\circ$	&	$\circ$	&	$\circ$	&	$\bullet$	&	$\bullet$	&	2048	&	$\bullet$	&	$\bullet$	&	$\bullet$	&	$\circ$	&	$\bullet$	\\
GoDaddy	&	2,359	&	15.0	&	B	&	$\circ$	&	$\bullet$	&	$\bullet$	&	$\circ$	&	$\circ$	&	$\bullet$	&	$\circ$	&	$\bullet$	&	$\circ$	&	$\circ$	&	$\circ$	&	$\circ$	&	$\circ$	&	$\bullet$	&	$\circ$	&	$\circ$	&	$\circ$	&	$\bullet$	&	$\bullet$	&	2048	&	$\bullet$	&	$\bullet$	&	$\bullet$	&	$\circ$	&	$\bullet$	\\
GoDaddy	&	968	&	6.2	&	B	&	$\circ$	&	$\bullet$	&	$\bullet$	&	$\bullet$	&	$\circ$	&	$\circ$	&	$\bullet$	&	$\bullet$	&	$\circ$	&	$\circ$	&	$\bullet$	&	$\circ$	&	$\circ$	&	$\circ$	&	$\circ$	&	$\circ$	&	$\circ$	&	$\bullet$	&	$\bullet$	&	2048	&	$\bullet$	&	$\bullet$	&	$\bullet$	&	$\circ$	&	$\bullet$	\\
GoDaddy	&	884	&	5.6	&	C	&	$\circ$	&	$\bullet$	&	$\bullet$	&	$\circ$	&	$\circ$	&	$\bullet$	&	$\bullet$	&	$\bullet$	&	$\bullet$	&	$\circ$	&	$\bullet$	&	$\circ$	&	$\circ$	&	$\bullet$	&	$\bullet$	&	$\bullet$	&	$\circ$	&	$\bullet$	&	$\bullet$	&	-	&	$\circ$	&	$\bullet$	&	$\bullet$	&	$\circ$	&	$\bullet$	\\
GoDaddy	&	630	&	4.0	&	B	&	$\circ$	&	$\bullet$	&	$\bullet$	&	$\bullet$	&	$\circ$	&	$\circ$	&	$\bullet$	&	$\bullet$	&	$\circ$	&	$\circ$	&	$\bullet$	&	$\circ$	&	$\circ$	&	$\bullet$	&	$\circ$	&	$\circ$	&	$\circ$	&	$\bullet$	&	$\bullet$	&	2048	&	$\bullet$	&	$\bullet$	&	$\bullet$	&	$\circ$	&	$\bullet$	\\
\midrule
Digital Ocean	&	1,570	&	11.4	&	B	&	$\circ$	&	$\bullet$	&	$\bullet$	&	$\bullet$	&	$\circ$	&	$\bullet$	&	$\bullet$	&	$\bullet$	&	$\circ$	&	$\circ$	&	$\bullet$	&	$\circ$	&	$\circ$	&	$\circ$	&	$\circ$	&	$\circ$	&	$\circ$	&	$\bullet$	&	$\bullet$	&	2048	&	$\bullet$	&	$\bullet$	&	$\bullet$	&	$\circ$	&	$\bullet$	\\
Digital Ocean	&	1,166	&	8.5	&	B	&	$\circ$	&	$\bullet$	&	$\bullet$	&	$\bullet$	&	$\circ$	&	$\bullet$	&	$\bullet$	&	$\bullet$	&	$\circ$	&	$\circ$	&	$\circ$	&	$\circ$	&	$\circ$	&	$\circ$	&	$\circ$	&	$\circ$	&	$\bullet$	&	$\bullet$	&	$\bullet$	&	2048	&	$\bullet$	&	$\bullet$	&	$\bullet$	&	$\circ$	&	$\bullet$	\\
Digital Ocean	&	995	&	7.2	&	B	&	$\bullet$	&	$\bullet$	&	$\bullet$	&	$\bullet$	&	$\circ$	&	$\bullet$	&	$\circ$	&	$\bullet$	&	$\circ$	&	$\circ$	&	$\circ$	&	$\circ$	&	$\bullet$	&	$\bullet$	&	$\circ$	&	$\bullet$	&	$\bullet$	&	$\bullet$	&	$\bullet$	&	2048	&	$\bullet$	&	$\bullet$	&	$\circ$	&	$\circ$	&	$\bullet$	\\
Digital Ocean	&	603	&	4.4	&	A	&	$\bullet$	&	$\bullet$	&	$\circ$	&	$\circ$	&	$\circ$	&	$\bullet$	&	$\circ$	&	$\bullet$	&	$\circ$	&	$\circ$	&	$\circ$	&	$\circ$	&	$\circ$	&	$\circ$	&	$\circ$	&	$\circ$	&	$\bullet$	&	$\bullet$	&	$\bullet$	&	-	&	$\circ$	&	$\bullet$	&	$\circ$	&	$\circ$	&	$\bullet$	\\
Digital Ocean	&	496	&	3.6	&	A	&	$\bullet$	&	$\bullet$	&	$\circ$	&	$\circ$	&	$\circ$	&	$\bullet$	&	$\circ$	&	$\bullet$	&	$\circ$	&	$\circ$	&	$\circ$	&	$\circ$	&	$\circ$	&	$\circ$	&	$\circ$	&	$\circ$	&	$\circ$	&	$\bullet$	&	$\bullet$	&	-	&	$\circ$	&	$\bullet$	&	$\bullet$	&	$\circ$	&	$\bullet$	\\
\midrule
SingleHop	&	7,446	&	70.9	&	B	&	$\bullet$	&	$\bullet$	&	$\bullet$	&	$\circ$	&	$\circ$	&	$\bullet$	&	$\bullet$	&	$\bullet$	&	$\circ$	&	$\circ$	&	$\circ$	&	$\circ$	&	$\bullet$	&	$\bullet$	&	$\circ$	&	$\circ$	&	$\bullet$	&	$\bullet$	&	$\bullet$	&	2048	&	$\bullet$	&	$\bullet$	&	$\bullet$	&	$\circ$	&	$\bullet$	\\
SingleHop	&	363	&	3.5	&	B	&	$\circ$	&	$\bullet$	&	$\circ$	&	$\circ$	&	$\circ$	&	$\circ$	&	$\bullet$	&	$\bullet$	&	$\circ$	&	$\circ$	&	$\circ$	&	$\circ$	&	$\circ$	&	$\circ$	&	$\circ$	&	$\circ$	&	$\circ$	&	$\bullet$	&	$\bullet$	&	-	&	$\circ$	&	$\bullet$	&	$\circ$	&	$\circ$	&	$\bullet$	\\
SingleHop	&	341	&	3.2	&	A	&	$\bullet$	&	$\bullet$	&	$\bullet$	&	$\bullet$	&	$\circ$	&	$\bullet$	&	$\bullet$	&	$\bullet$	&	$\circ$	&	$\circ$	&	$\bullet$	&	$\circ$	&	$\circ$	&	$\circ$	&	$\circ$	&	$\circ$	&	$\bullet$	&	$\bullet$	&	$\bullet$	&	-	&	$\circ$	&	$\bullet$	&	$\bullet$	&	$\circ$	&	$\bullet$	\\
SingleHop	&	196	&	1.9	&	B	&	$\circ$	&	$\bullet$	&	$\bullet$	&	$\bullet$	&	$\circ$	&	$\circ$	&	$\bullet$	&	$\bullet$	&	$\circ$	&	$\circ$	&	$\bullet$	&	$\circ$	&	$\circ$	&	$\circ$	&	$\circ$	&	$\circ$	&	$\circ$	&	$\bullet$	&	$\bullet$	&	2048	&	$\bullet$	&	$\bullet$	&	$\bullet$	&	$\circ$	&	$\bullet$	\\
SingleHop	&	149	&	1.4	&	A	&	$\circ$	&	$\bullet$	&	$\circ$	&	$\circ$	&	$\circ$	&	$\bullet$	&	$\bullet$	&	$\bullet$	&	$\circ$	&	$\circ$	&	$\circ$	&	$\circ$	&	$\circ$	&	$\circ$	&	$\circ$	&	$\circ$	&	$\bullet$	&	$\bullet$	&	$\bullet$	&	-	&	$\circ$	&	$\bullet$	&	$\circ$	&	$\circ$	&	$\bullet$	\\
\bottomrule
\end{tabular*}

%% file: tables/appendix_stats.tex
\begin{tabular}{llllll|llllllll}
\toprule
	& \multicolumn{5}{c}{Tranco Top 1 Million} & \multicolumn{5}{c}{Certificate
	Transparency Sample} \\ \cmidrule(r){2-6} \cmidrule(r){7-11}
TLS Configuration & All & Apache & Nginx & IIS & Other & All & Apache & Nginx & IIS & Other\\
\midrule
Version Support\\
\quad TLS v13 	&	0.28 & 0.04	&	0.15	&	0.01	&	0.23	&	0.12	&	0.02	&	0.11	&	0.01	&	0.21 \\
\quad TLS v12	&	0.99	&	0.99	&	0.99	&	0.91	&	0.99	&	0.99	&	1	&	1	&	0.86	&	1 \\
\quad TLS v11	&	0.81	&	0.78	&	0.85	&	0.75	&	0.7	&	0.75	&	0.74	&	0.87	&	0.78	&	0.69 \\
\quad TLS v10	&	0.68	&	0.64	&	0.64	&	0.76	&	0.53	&	0.68	&	0.67	&	0.71	&	0.88	&	0.62 \\
\quad SSL v30	&	0.04	&	0.01	&	0.03	&	0.21	&	0.02	&	0.02	&	0.01	&	0.01	&	0.29	&	0.01 \\
\quad SSL v20	&	0.01	&	0	&	0	&	0.06	&	0	&	0	&	0	&	0	&	0.11	&	0 \\
\quad HTTP/2	&	0.57	&	0.26	&	0.66	&	0.23	&	0.63	&	0.56	&	0.42	&	0.75	&	0.24	&	0.64 \\

Compression \\
\quad	 TLS compression 	&	0	&	0	&	0	&	0	&	0	&	0	&	0	&	0	&	0	&	0 \\
\quad	HTTPS compression gzip	&	0.42	&	0.38	&	0.48	&	0.29	&	0.36	&	0.5	&	0.47	&	0.51	&	0.36	&	0.57 \\
\quad	HTTP2 compression gzip	&	0.28	&	0.14	&	0.32	&	0.06	&	0.27	&	0.3	&	0.23	&	0.38	&	0.08	&	0.37 \\

Extensions \\
\quad server name	&	0.61	&	0.89	&	0.86	&	0.45	&	0.52	&	0.75	&	0.97	&	0.88	&	0.47	&	0.44 \\
\quad status request	&	0.41	&	0.30	&	0.19	&	0.45	&	0.36	&	0.42	&	0.42	&	0.13	&	0.4	&	0.53 \\
\quad signed certificate timestamp	&	0	&	0	&	0	&	0	&	0	&	0	&	0	&	0	&	0	&	0 \\
\quad heartbeat	&	0.45	&	0.84	&	0.6	&	0.03	&	0.24	&	0.65	&	0.93	&	0.57	&	0.03	&	0.38 \\
\quad extended master secret &	0.45	&	0.09	&	0.32	&	0.84	&	0.51	&	0.28	&	0.05	&	0.39	&	0.85	&	0.47 \\
\quad session ticket	&	0.83	&	0.89	&	0.87	&	0.06	&	0.73	&	0.88	&	0.94	&	0.83	&	0.06	&	0.87 \\
\quad renegotiation info	&	0.99	&	0.99	&	1	&	0.99	&	0.99	&	0.99	&	0.99	&	1	&	0.99	&	0.99 \\
\quad server preference support	&	0.63	&	0.59	&	0.91	&	0.99	&	0.9	&	0.67	&	0.54	&	0.93	&	0.98	&	0.77 \\
\quad	session id resumption	&	0.82	&	0.91	&	0.68	&	0.86	&	0.64	&	0.83	&	0.95	&	0.73	&	0.9	&	0.69 \\
\quad alpn	&	0.77	&	0.07	&	0.81	&	0.26	&	0.78	&	0.88	&	0.89	&	0.9	&	0.27	&	0.88 \\

Ciphers \\
\quad	 RC4	&	0.08	&	0.14	&	0.01	&	0.41	&	0.06	&	0.08	&	0.08	&	0.01	&	0.57	&	0.09 \\
\quad	 DES	&	0	&	0.01	&	0	&	0.01	&	0	&	0	&	0	&	0	&	0.01	&	0 \\
\quad	 3DES	&	0.30	&	0.43	&	0.24	&	0.67	&	0.35	&	0.32	&	0.31	&	0.26	&	0.78	&	0.37\\
\quad	 EXPORT	&	0	&	0	&	0	&	0	&	0	&	0	&	0	&	0	&	0	&	0\\
\quad	 SEED	&	0.07	&	0.15	&	0.03	&	0.01	&	0.05	&	0.07	&	0.1	&	0.02	&	0.01	&	0.07\\
\quad	 CAMELLIA	&	0.29	&	0.43	&	0.46	&	0.04	&	0.25	&	0.27	&	0.26	&	0.41	&	0.03	&	0.24\\
\quad	 ARIA	&	0.05	&	0.02	&	0.15	&	0	&	0.01	&	0.05	&	0.01	&	0.24	&	0	&	0.01\\
\quad	 IDEA	&	0.04	&	0.11	&	0.01	&	0	&	0.04	&	0.06	&	0.08	&	0.01	&	0.01	&	0.06\\
\quad AES & 1 & 1 & 1 & 1 & 1 & 1 & 1 & 1 & 1 & 1  \\
\quad MD5	&	0.07	&	0.11	&	0.01	&	0.4	&	0.06	&	0.07	&	0.07	&	0.01	&	0.56	&	0.06\\ 
\quad SHA1   & 0.94 & 0.86 & 0.96 & 0.99 & 0.96 & 0.87 & 0.79 & 0.96 & 0.99 & 0.87  \\
\quad SHA2   & 1.0 & 1.0 & 1.0 & 1.0 & 1.0 	& 1.0 & 1.0 & 1.0 & 1.0 & 1.0  \\

Curves \\
\quad	  sect283k1/r1/s	&	0.25	&	0.51	&	0.26	&	0.02	&	0.16	&	0.53	&	0.8	&	0.29	&	0.01	&	0.33 \\
\quad	  sect409k1/r1/s	&	0.25	&	0.51	&	0.26	&	0.02	&	0.16	&	0.53	&	0.8	&	0.29	&	0.01	&	0.33 \\
\quad	  sect571k1/r1/s	&	0.25	&	0.51	&	0.26	&	0.02	&	0.16	&	0.53	&	0.8	&	0.29	&	0.01	&	0.33 \\
\\quad	  secp224r1/s	&	0.21	&	0.01	&	0.01	&	0.02	&	0.03	&	0.03	&	0	&	0	&	0.01	&	0.01 \\
\quad	  secp256k1	&	0.31	&	0.59	&	0.38	&	0.03	&	0.19	&	0.56	&	0.82	&	0.39	&	0.02	&	0.34 \\
\quad	  secp256r1	&	0.96	&	0.97	&	0.9	&	0.97	&	0.98	&	0.97	&	0.99	&	0.91	&	0.98	&	0.98 \\
\quad	  secp256s	&	0.31	&	0.59	&	0.38	&	0.03	&	0.19	&	0.56	&	0.82	&	0.39	&	0.02	&	0.34 \\
\quad	  secp384r1	&	0.80	&	0.71	&	0.75	&	0.93	&	0.8	&	0.86	&	0.87	&	0.82	&	0.9	&	0.86 \\
\quad	  secp521r1	&	0.66	&	0.68	&	0.63	&	0.26	&	0.43	&	0.73	&	0.86	&	0.71	&	0.17	&	0.54 \\
\quad	  brainpool 256r1, 384r1, 512r1	&	0.25	&	0.51	&	0.26	&	0.02	&	0.16	&	0.53	&	0.8	&	0.3	&	0.01	&	0.33 \\
\quad	  x25519	&	0.39	&	0.08	&	0.28	&	0.21	&	0.47	&	0.27	&	0.04	&	0.35	&	0.21	&	0.52 \\
\quad	  x448	&	0.08	&	0.05	&	0.16	&	0.01	&	0.12	&	0.08	&	0.03	&	0.27	&	0	&	0.06 \\

Key Exchange \\
\quad	RSA	&	0.68	&	0.71	&	0.78	&	0.95	&	0.81	&	0.56	&	0.48	&	0.65	&	0.96	&	0.64 \\
\quad	DHE	&	0.44	&	0.71	&	0.54	&	0.49	&	0.31	&	0.55	&	0.7	&	0.58	&	0.54	&	0.35 \\
\quad ECHDE	&	0.99	&	0.97	&	0.99	&	0.98	&	0.99	&	0.99	&	0.99	&	1	&	0.98	&	1 \\
\quad	Common DH group	&	0.32	&	0.66	&	0.23	&	0.46	&	0.18	&	0.46	&	0.68	&	0.25	&	0.51	&	0.29 \\
\quad	DH group size 0-1024	&	0	&	0	&	0	&	0	&	0	&	0	&	0	&	0	&	0	&	0 \\
\quad	DH group size 1024-2048	&	0.05	&	0.03	&	0.07	&	0.3	&	0.05	&	0.02	&	0.01	&	0.03	&	0.36	&	0.01 \\
\quad	DH group size 2048+	&	0.39	&	0.67	&	0.47	&	0.2	&	0.27	&	0.53	&	0.69	&	0.54	&	0.17	&	0.34 \\
\bottomrule
\end{tabular}

%% file: tables/appendix_model.tex
\begin{tabular}{lllll|llll|llll}
\toprule
& \multicolumn{4}{c}{F/C} & \multicolumn{4}{c}{B} & \multicolumn{4}{c}{A} \\ \cmidrule(r){2-5} 
\cmidrule(r){6-9} 
\cmidrule(r){10-13}
	&	Marg. Odds	&	t value	&	Pr($>\vert$t$\vert$)	&		&	Marg. Odds	&	t value	&	Pr($>\vert$t$\vert$)	&		&	Marg. Odds	&	t value	&	Pr($>\vert$t$\vert$)	&	\\
\midrule			
servernginx	&	0.94	&	-46.67	&	< 0.00	&	***	&	0.99	&	-35.67	&	< 0.00	&	***	&	1.07	&	45.74	&	< 0.00	&	***	\\
serverother	&	0.87	&	-95.46	&	< 0.00	&	***	&	0.96	&	-57.49	&	< 0.00	&	***	&	1.2	&	86.55	&	< 0.00	&	***	\\
servermicrosoft-iis	&	1.1	&	37.19	&	< 0.00	&	***	&	1.01	&	21.86	&	< 0.00	&	***	&	0.9	&	-40.93	&	< 0.00	&	***	\\
serverlitespeed	&	0.73	&	-261.56	&	< 0.00	&	***	&	0.81	&	-243.73	&	< 0.00	&	***	&	1.71	&	295.08	&	< 0.00	&	***	\\
serveropenresty	&	0.88	&	-41.38	&	< 0.00	&	***	&	0.95	&	-25.16	&	< 0.00	&	***	&	1.19	&	35.7	&	< 0.00	&	***	\\
servercpanel	&	1.1	&	19.45	&	< 0.00	&	***	&	1	&	12.84	&	< 0.00	&	***	&	0.9	&	-21.56	&	< 0.00	&	***	\\
serverbigip	&	0.87	&	-39.52	&	< 0.00	&	***	&	0.95	&	-23.92	&	< 0.00	&	***	&	1.2	&	33.94	&	< 0.00	&	***	\\
servercloudfront	&	0.72	&	-491.91	&	< 0.00	&	***	&	0.79	&	-910.96	&	< 0.00	&	***	&	1.77	&	688.48	&	< 0.00	&	***	\\
servervarnish	&	0.83	&	-50.58	&	< 0.00	&	***	&	0.92	&	-27.87	&	< 0.00	&	***	&	1.32	&	40.46	&	< 0.00	&	***	\\
serverats	&	0.92	&	-15.67	&	< 0.00	&	***	&	0.98	&	-10.66	&	< 0.00	&	***	&	1.11	&	14.24	&	< 0.00	&	***	\\
serverawselb/2.0	&	0.73	&	-236.95	&	< 0.00	&	***	&	0.79	&	-194.89	&	< 0.00	&	***	&	1.72	&	239.89	&	< 0.00	&	***	\\
serversquarespace	&	0.98	&	-3.32	&	0.0009	&	***	&	1	&	-2.86	&	0.0043	&	**	&	1.03	&	3.24	&	0.0012	&	**	\\
serverakamai	&	0.75	&	-131.25	&	< 0.00	&	***	&	0.81	&	-69.02	&	< 0.00	&	***	&	1.64	&	98.22	&	< 0.00	&	***	\\
serverghs	&	0.97	&	-3.2	&	0.0014	&	**	&	0.99	&	-2.7	&	0.0069	&	**	&	1.03	&	3.11	&	0.0019	&	**	\\
median rankQ2	&	1	&	1.16	&	0.2479	&		&	1	&	1.17	&	0.2437	&		&	1	&	-1.16	&	0.2473	&		\\
median rankQ3	&	1	&	0.73	&	0.4666	&		&	1	&	0.75	&	0.4547	&		&	1	&	-0.73	&	0.465	&		\\
median rankQ4	&	1	&	0.78	&	0.4373	&		&	1	&	0.78	&	0.4345	&		&	1	&	-0.78	&	0.4369	&		\\
median rankQ5	&	1	&	0.04	&	0.9715	&		&	1	&	0.04	&	0.9715	&		&	1	&	-0.04	&	0.9715	&		\\
sd rankQ2	&	1.01	&	4.36	&	< 0.00	&	***	&	1	&	4.49	&	< 0.00	&	***	&	0.99	&	-4.38	&	0	&	***	\\
sd rankQ3	&	1.01	&	3.5	&	0.0005	&	***	&	1	&	3.64	&	0.0003	&	***	&	0.99	&	-3.52	&	0.0004	&	***	\\
sd rankQ4	&	1	&	1.71	&	0.0881	&		&	1	&	1.76	&	0.0789	&		&	1	&	-1.71	&	0.0868	&		\\
sd rankQ5	&	1	&	1.45	&	0.147	&		&	1	&	1.47	&	0.1424	&		&	1	&	-1.45	&	0.1464	&		\\
tld org	&	0.99	&	-4.98	&	< 0.00	&	***	&	1	&	-4.65	&	< 0.00	&	***	&	1.01	&	4.93	&	0	&	***	\\
tld net	&	1.01	&	4.12	&	< 0.00	&	***	&	1	&	4.55	&	< 0.00	&	***	&	0.99	&	-4.18	&	0	&	***	\\
tld ru	&	1	&	-0.38	&	0.7008	&		&	1	&	-0.38	&	0.7033	&		&	1	&	0.38	&	0.7011	&		\\
tld de	&	0.98	&	-6.01	&	< 0.00	&	***	&	1	&	-5.23	&	< 0.00	&	***	&	1.03	&	5.87	&	0	&	***	\\
tld uk	&	0.99	&	-3.11	&	0.0019	&	**	&	1	&	-2.84	&	0.0045	&	**	&	1.02	&	3.07	&	0.0022	&	**	\\
tld jp	&	1.05	&	9.74	&	< 0.00	&	***	&	1.01	&	20.13	&	< 0.00	&	***	&	0.94	&	-10.29	&	< 0.00	&	***	\\
tld br	&	1.01	&	1.94	&	0.0522	&	0	&	1	&	2.12	&	0.0341	&	*	&	0.99	&	-1.96	&	0.0497	&	*	\\
tld in	&	1.01	&	2.3	&	0.0217	&	*	&	1	&	2.57	&	0.0102	&	*	&	0.99	&	-2.33	&	0.0199	&	*	\\
tld ae	&	1.02	&	2.63	&	0.0087	&	**	&	1	&	3.02	&	0.0025	&	**	&	0.98	&	-2.67	&	0.0076	&	**	\\
tld nl	&	0.97	&	-5.81	&	< 0.00	&	***	&	0.99	&	-4.79	&	< 0.00	&	***	&	1.04	&	5.61	&	< 0.00	&	***	\\
tld fr	&	1.01	&	1.98	&	0.0481	&	*	&	1	&	2.19	&	0.0282	&	*	&	0.99	&	-2	&	0.0453	&	*	\\
tld au	&	0.97	&	-4.24	&	< 0.00	&	***	&	0.99	&	-3.58	&	0.0004	&	***	&	1.03	&	4.11	&	< 0.00	&	***	\\
tld pl	&	1	&	-0.26	&	0.7936	&		&	1	&	-0.26	&	0.7962	&		&	1	&	0.26	&	0.794	&		\\
tld ir	&	1.03	&	3.82	&	0.0001	&	***	&	1	&	5.2	&	< 0.00	&	***	&	0.97	&	-3.94	&	< 0.00	&	***	\\
tld it	&	1.03	&	4.74	&	< 0.00	&	***	&	1	&	6.77	&	< 0.00	&	***	&	0.97	&	-4.9	&	< 0.00	&	***	\\
tld info	&	1	&	-0.47	&	0.6362	&		&	1	&	-0.46	&	0.6447	&		&	1	&	0.47	&	0.6374	&		\\
tld ca	&	0.99	&	-1	&	0.318	&		&	1	&	-0.95	&	0.3432	&		&	1.01	&	0.99	&	0.3217	&		\\
tld co	&	0.97	&	-4.26	&	< 0.00	&	***	&	0.99	&	-3.53	&	0.0004	&	***	&	1.04	&	4.12	&	0	&	***	\\
tld ua	&	0.96	&	-5.37	&	< 0.00	&	***	&	0.99	&	-4.26	&	< 0.00	&	***	&	1.05	&	5.14	&	< 0.00	&	***	\\
tld edu	&	1.01	&	0.73	&	0.4635	&		&	1	&	0.77	&	0.4411	&		&	0.99	&	-0.74	&	0.4606	&		\\
tld io	&	0.94	&	-8.82	&	< 0.00	&	***	&	0.98	&	-6.31	&	< 0.00	&	***	&	1.09	&	8.16	&	< 0.00	&	***	\\
tld es	&	1.02	&	2.51	&	0.012	&	*	&	1	&	3.11	&	0.0019	&	**	&	0.98	&	-2.57	&	0.0101	&	*	\\
tld vn	&	1.05	&	4.95	&	< 0.00	&	***	&	1	&	10.82	&	< 0.00	&	***	&	0.95	&	-5.22	&	< 0.00	&	***	\\
tld cz	&	1.01	&	1.44	&	0.151	&		&	1	&	1.61	&	0.1076	&		&	0.99	&	-1.46	&	0.1454	&		\\
tld other	&	1.01	&	4.63	&	< 0.00	&	***	&	1	&	4.85	&	< 0.00	&	***	&	0.99	&	-4.66	&	< 0.00	&	***	\\
\bottomrule
\end{tabular}